\documentclass[sensors,review,submit,moreauthors,pdftex]{Definitions/mdpi} 

\usepackage{url}
\usepackage{hyperref}
\usepackage{pifont}
\usepackage{dirtytalk}

\newcommand{\cmark}{\ding{51}}%
\newcommand{\xmark}{\ding{55}}%
\newcommand{\p}{{\cmark}}
\newcommand{\n}{{\xmark}}
\newcommand{\co}{\textbf{\large{--}}}
\newcommand{\sectorsSep}{Healthcare; Smart Homes; Smart Buildings; Smart Cities; Transportation; Industrial IoT}
\newcommand{\Sectors}{Healthcare, Smart Homes,  Smart Buildings, Smart Cities, Transportation and Industrial IoT}
\newcommand{\sectors}{healthcare, smart homes,  smart buildings, smart cities, transportation and industrial IoT}

\newcommand{\change}[1]{{\color{black}{#1}}}

\renewcommand{\hl}[1]{{\color{black}{#1}}}

\preto{\abstractkeywords}{\nolinenumbers}

\firstpage{1} 
\pubvolume{21}
\issuenum{11}
\articlenumber{3838}
\pubyear{2021}
\copyrightyear{2021}
\datereceived{} 
\dateaccepted{} 
\datepublished{} 
\hreflink{https://doi.org/10.3390/s21113838} 

\Title{Impact of COVID-19 on IoT Adoption in \Sectors}

\TitleCitation{Impact of COVID-19 on IoT Adoption in \Sectors}

\Author{Muhammad Umair $^{1}$\orcidA{}, Muhammad Aamir Cheema $^2$$^,$$^*$\orcidB{}, Omer Cheema $^{3}$, Huan Li $^{4}$\orcidD{} and Hua Lu $^{5}$\orcidE{}}

\AuthorNames{Firstname Lastname, Firstname Lastname and Firstname Lastname}

\AuthorCitation{Umair, M.; Cheema, M. A.; Cheema, O.; Li, H.; Lu, H.}

\address{%
$^{1}$ \quad Department of Electrical, Electronics and Telecommunication Engineering, New Campus, University of Engineering and Technology, Lahore, Punjab, Pakistan; m.umair@uet.edu.pk\\
$^{2}$ \quad Faculty of Information Technology, Monash University, Wellington Rd, Clayton VIC 3800, Australia; aamir.cheema@monash.edu\\
$^{3}$ \quad IoT Wi-Fi business unit, Dialog Semiconductor, Green Park Reading RG2 6GP, United Kingdom; omer.cheema@diasemi.com\\
$^{4}$ \quad Department of Computer Science, Aalborg University, Fredrik Bajers Vej 7K, 9220 Aalborg Øst, Denmark; lihuan@cs.aau.dk\\
$^{5}$ \quad Department of People and Technology, Roskilde University, 1, 4000 Roskilde, Denmark; luhua@ruc.dk}

\corres{Correspondence: aamir.cheema@monash.edu; Tel.: +61-3-9905-9662}

\abstract{COVID-19 has disrupted normal life and has enforced a substantial change in the policies, priorities and activities of individuals, organisations and governments. These changes are proving to be a catalyst for technology and innovation. In this paper, we discuss the pandemic's potential impact on the adoption of the Internet of Things (IoT) in various broad sectors namely \sectors. Our perspective and forecast of this impact on IoT adoption is based on a thorough research literature review, a careful examination of reports from leading consulting firms and interactions with several industry experts. For each of these sectors, we also provide the details of notable IoT initiatives taken in wake of COVID-19. We also highlight the challenges that need to be addressed and important research directions that will facilitate accelerated IoT adoption.}


\begin{document}

\section{Introduction}\label{sec:intro}

The Internet of Things (IoT) is envisioned as a network of billions of devices that can sense, communicate and share data which can then be analyzed to unlock a wealth of intelligence useful for planning, management, and decision making~\cite{patel2016internet}. IoT promises huge benefits for a variety of domains such as healthcare, manufacturing, agriculture, telecommunication and transportation.   
Despite the popularity of the IoT concept and its promised benefits, its adoption  has been significantly slower than expected~\cite{patel2017s,farhan2017survey}. Some of the major reasons behind this include: 1) security, privacy, policy, and trust issues~\cite{farhan2017survey,lin2016research,jalali2019internet}; 2) organisational inertia, long capital cycles and shortage of specialist workforce needed to successfully implement IoT~\cite{patel2017s,jalali2019internet}; and 3) lack of convincing use cases with clear return on investment (ROI)
in some sectors.

COVID-19 has impacted all walks of life so much that we may never return to the old normal. This pandemic  is proving to be a catalyst for digital transformation because COVID-19 has created or expanded applications and use cases of digital technologies~\cite{renu2021technological,eluneven,anderson2020business}. It has also forced governments, organisations and individuals to change/adapt their priorities, their views on societal/ethical issues, and the way they operate.  In many cases, this has  addressed or mitigated many of the above mentioned reasons behind the slower-than-expected adoption of IoT across many verticals. For example, governments have invested huge amount of resources in IoT and other technologies to combat COVID-19. Lifestyle changes brought about by COVID-19 such as work/study from home have also provided new use cases of IoT with clear ROI such as remote asset control, workforce tracking and remote employee collaboration~\cite{McKinsey1}. Consequently, many organisations have increased investment in IoT and the pace of their IoT projects~\cite{bitfidner,ComputerWeekly}.
Also, the fight against COVID-19 has resulted in less strict stance on privacy issues, higher trust in technology and fast-tracked approval procedures.
This is also paving the way for accelerated adoption of IoT in various verticals. Furthermore, regulatory changes such as stricter cleaning and tracking requirements for businesses are also accelerating the IoT adoption in smart buildings. 

In this paper, we discuss the potential impact of COVID-19 on the adoption of IoT in different sectors namely \sectors.
To this end, we have extensively analyzed recent relevant research literature and examined existing reports from Gartner, Yole, McKinsey and other consulting firms. 
We find that COVID-19 has not necessarily positively impacted the adoption of IoT in all sectors, at least in the short-term. For example, mainly due to the recession, most automotive companies are not in a position to invest resources in IoT initiatives. Therefore,  we discuss both the short-term and mid- to long-term impact of COVID-19 on IoT adoption in these sectors. Furthermore, for each of these sectors, we also discuss new initiatives that are being taken, challenges that need to be addressed, and important research directions that will facilitate IoT adoption.

To the best of our knowledge, this is the first work that details the impact of COVID-19 on the adoption of IoT in different sectors.
Since the beginning of COVID-19, numerous papers have been published discussing IoT and COVID-19 (see \autoref{sec:related}). However, our work is significantly different in the following major aspects. Firstly, an overwhelming majority of the existing works exclusively focus on IoT in healthcare sector. This is not surprising because COVID-19 is a health crisis. However, COVID-19 has impacted almost every aspect of our life and we offer the first report about its impact on IoT in a variety of important sectors including healthcare. Secondly, the existing studies mostly focus on how IoT and other digital technologies can be used (or are being used) to combat COVID-19. On the other hand, we do not only provide an overview of recent IoT initiatives in wake of COVID-19 across different sectors but also discuss the impact of COVID-19 on the adoption of IoT in these sectors. In other words, the existing works focus on how IoT can or should impact COVID-19, whereas we also focus on how COVID-19 is impacting (or expected to impact) the IoT sectors.
Finally, many of the existing works only discuss existing and/or potential applications of IoT for COVID-19. In this paper, we also discuss the impact of COVID-19 on IoT implementations and research challenges that need to be tackled to facilitate IoT adoption in different sectors.

 The rest of the paper is organized as follows.
 In \autoref{sec:related}, we provide an overview of the existing studies that discuss COVID-19 and digital technologies, particularly focusing on IoT. In \autoref{sec:impact}, we detail the impact of COVID-19 on different IoT sectors namely \sectors. We also discuss notable recent initiatives in each of these sectors taken in the wake of COVID-19.
The challenges that must be addressed and important research directions to facilitate accelerated IoT adoption are discussed in \autoref{sec:challenges}. \autoref{sec:conclusion} concludes the paper.
\section{Related Work}\label{sec:related}

In the past year or so, numerous studies have  discussed key potentials of IoT and other digital technologies in the fight against COVID-19 or future pandemics. \citet{elansary2021future} present a survey of IoT-based solutions used to fight COVID-19. \citet{brem2021implications} analyse the effect of this pandemic on various technologies and discuss their social impacts.  A detailed review of digital health solutions used in countries with high COVID-19 cases is presented in~\cite{kalhori2021digital}. A consensus of Chinese experts on IoT-aided diagnosis of COVID-19 and its treatment is presented in~\cite{IEEEhowto:Bai}. Impacts of IoT implementation in healthcare in terms of cost, time and efficiency are enlisted in \cite{IoTHealthcare}. Applications of IoT, Big Data, Artificial Intelligence (AI) and Blockchain in mitigating the impact of COVID-19 are explored in~\cite{DigitalTechCOVID}. \citet{IoTCOVID} propose some IoT applications that can be useful to combat COVID-19.  \citet{Industry4.0Covid} discuss how different industry 4.0 technologies (e.g., AI, IoT, Virtual Reality, etc.) can help reduce the spread of disease. Applications of AI for COVID-19 have been proposed in \cite{AICovid}. A comprehensive review of the COVID-19 pandemic and the role of IoT, drones, AI, blockchain, and 5G in managing its impact is explored in \cite{IoTAICovid}. \citet{ContactTracing} argue that contact tracing should be the responsibility of facilities and propose a contact tracing architecture which is fully automated and does not depend on user cooperation. \citet{IoTandCovid} discuss several IoT healthcare applications during three main phases: early diagnosis, quarantine time, and after recovery. A recent survey~\cite{nayakintelligent} discusses the use of Machine Learning (ML), AI and other intelligent approaches for the prognosis of COVID-19.

\citet{dong2020iot} have compiled potential IoT-based solutions to combat COVID-19. They present a detailed study on the capabilities of existing IoT systems at different layers such as perception layer, network layer, fog layer and cloud layer. Moreover, they also discuss applications of IoT in diagnosing symptoms of COVID-19. A four-layered architecture based on IoT and Blockchain technologies has been proposed in \cite{alam2020internet} to help fight against COVID-19. The Blockchain-based method is proposed to ensure privacy and security of physiological information shared among IoT nodes. It also enlists various applications that have been developed for detecting and tracing potential COVID-19 patients. The role of IoT in existing digital healthcare infrastructure has been discussed in \cite{kelly2020internet}. It also debates on the implications of data generated through IoT enabled healthcare infrastructure on the decisions made by policy makers. Moreover, existing enablers and barriers in adopting IoT-based healthcare have also been enlisted.

A detailed survey on the contributions of IoT in healthcare in response to COVID-19 is provided in \cite{ndiaye2020iot}. This is a detailed study enlisting recent developments in Healthcare IoT (HIoT). It also outlines comparison of different implementation strategies for IoT systems before and during this pandemic.  \citet{golinelli2020adoption} present a survey enlisting early efforts for the adoption of digital technologies in healthcare to fight against COVID-19, considering different categories such as diagnosis, prevention and surveillance.
\citet{chang2020artificial} emphasize the dire need of using AI techniques to combat future pandemics. They also highlight the limitations of the existing AI-based approaches towards eradication of the pandemic. Finally, they conclude that there is a need to use data science in global health  to produce better predictions helpful for policy makers.  A comparison of adoption of digital technologies in some specific regions of the world is provided in \cite{eluneven}. 
The paper highlights that COVID-19 has acted as a catalyst for adoption of e-Health, e-Education and e-Commerce. \citet{abir2020building} present an analysis of how AI and IoT can be potentially used to fight against the COVID-19 pandemic.

As detailed in \autoref{sec:intro}, there are some major differences between our paper and the above mentioned studies. Firstly, we have a broader focus and discuss IoT not only in healthcare but also in other important sectors such as smart homes, smart buildings, smart cities, transportation and industrial IoT. Secondly, while most of the existing studies focus on how IoT can impact COVID-19 and future pandemics, our focus is on how  COVID-19 is affecting (and expected to affect) the adoption of IoT in different sectors. Finally, we discuss new IoT initiatives taken in different sectors, and present the challenges that need to be addressed to facilitate IoT adoption.

\end{paracol}
\begin{figure}[htbp]
\widefigure
\includegraphics[width=\textwidth]{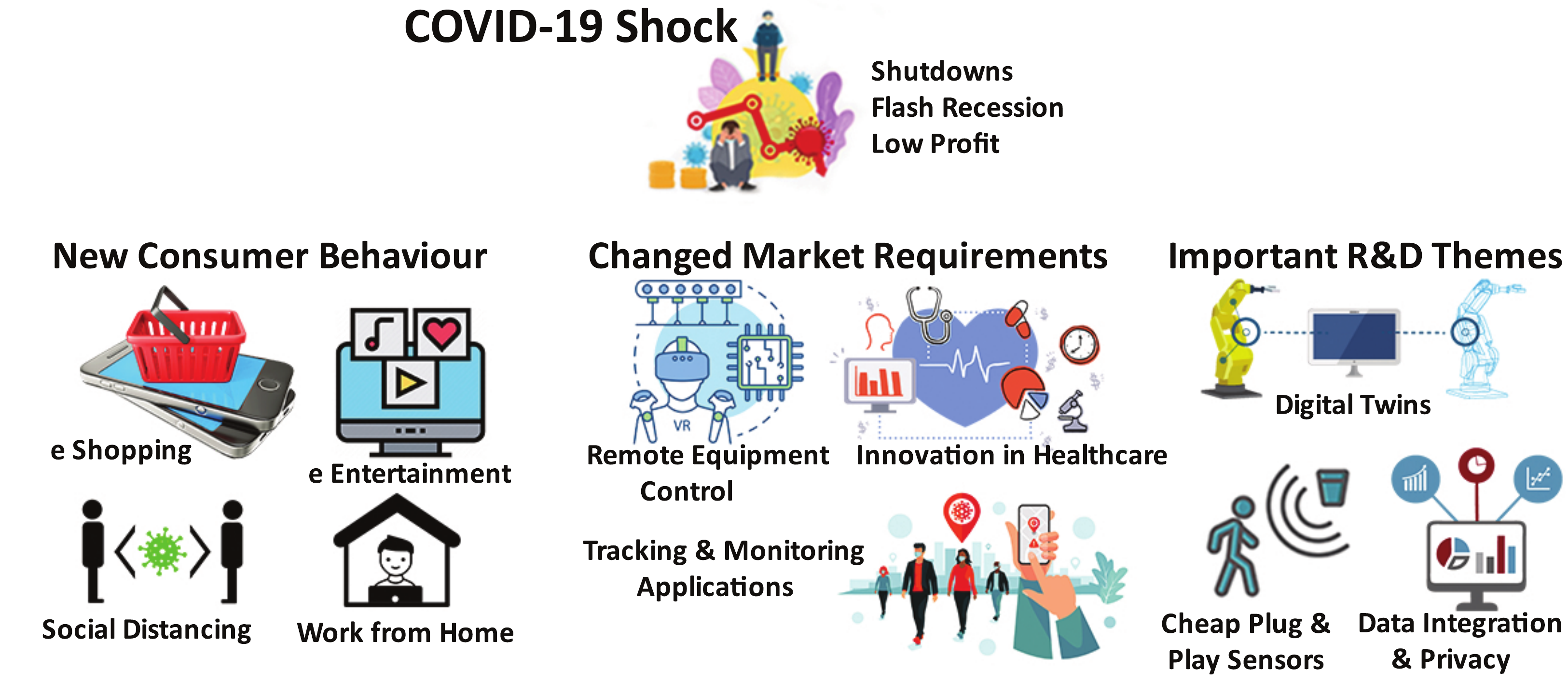}
\caption{An example of COVID-19's Impact on IoT Research and Development}
\label{fig_rd-themes}
\end{figure}

\begin{paracol}{2}
\switchcolumn

\section{Impact of COVID-19 on IoT and New Initiatives}\label{sec:impact}

COVID-19, as a macroeconomic shock, has  impacted not only individual behaviours but has also enforced governments and organisations to change their policies and priorities. This is fueling the adoption of IoT and other technologies in many areas especially in healthcare and smart cities. On the other hand, financial stress brought by the global recession has adversely impacted the technology adoption in the short-term in some sectors such as  transportation. However, in the mid- to long-term when financial stress is eased, this pandemic is expected to accelerate IoT adoption in a broader range of sectors.

The changes in consumer behavior, government policies, and organisation priorities open a range of opportunities to adopt IoT for innovative solutions to prepare for the new normal. \change{Many of the changes in this new normal such as remote work and telehealth etc. are expected to stay even after we get over the pandemic. For example, Pew Research Center recently conducted a study~\cite{PewResearchNewNormal} where more than 900 innovators, developers, business and policy leaders, researchers and activists were asked to consider life in 2025 in the wake of COVID-19. Most of these experts believe that the future will be much more technology driven as a large population comes to rely more on ``tele-everything'' such as remote  work, education, health care, daily commercial transactions and essential social interactions etc.} Therefore, researchers and developers need to come up with new products and services that align with the changed consumer behaviors.  The new product and service requirements will drive the organisations to reconsider their R\&D budgets in order to align with new research themes. 
\autoref{fig_rd-themes} shows an example of how COVID-19 is affecting IoT research and development. 

In this section, we discuss the potential impact of COVID-19 on IoT adoption in \sectors. We also discuss the most notable recent initiatives taken in each of these sectors in the wake of COVID-19.
To this end, we conducted an extensive literature review by searching scholarly databases using the terms ``IoT'', ``COVID-19'' and other equivalent terms (e.g., coronavirus, SARS-CoV-2 and internet of things etc.). We also set alerts on Google Scholar so that we do not miss recently published papers. Furthermore, we carefully examined reports from leading consulting firms such as Gartners, Yole and McKinsey as well as relevant articles/reports from reputable online sources.
Additionally, we also interviewed people from various companies and organisations including Apple, Amazon, Facebook (Oculus), Google (Glasses), Ford, BMW, Tesla, Honeywell, Stanford, Technion, and Huawei etc.
Questions were asked regarding the short-term and long-term impact of COVID-19 on the interviewee's industry, what was the industry doing to avoid or mitigate the crisis, how was the pandemic impacting their technology roadmaps, what new applications of IoT were being considered in the wake of COVID-19, and how had market needs changed etc.
Due to the privacy and policy requirements of these companies, in this paper, we are unable to give details of anything specific discussed during the interviews. However, while our perspective and forecast on the impact of COVID-19 on different IoT sectors presented in this paper is often inspired by our interactions with these experts, we support the claims by citing relevant research papers, reports or articles wherever necessary.

We remark that some of the sectors we discuss in this paper overlap with each other (e.g., smart transportation and smart homes are often considered to be encompassed by smart cities). In each subsection, we focus on themes that we believe are the most closely related to the sector, e.g., in-cabin air quality analytics for public transport is discussed in transportation and not in smart cities.

 \subsection{Healthcare} 
COVID-19 is a healthcare crisis and hence the immediate impact on healthcare is more obvious. A study from Juniper Research found that IoT platform revenue is expected to increase by around 20\% in 2020, reaching \$66 billion in 2020 from \$55 billion in 2019, mainly due to its accelerated adoption in healthcare sector~\cite{Juniper_Adoption}. 
Three main themes are accelerating IoT adoption in healthcare sector as detailed next.

\subsubsection{Wearable Devices}
While wearables, such as smart watches, smart bands and finger rings, have been around for several years, COVID-19 has triggered a huge increase in their demand.
Wearables can play vital roles in fighting against COVID-19 and other future pandemics~\cite{sawyer2020wearable}.  For example, the data from wearable devices can be used to alert the users when changes in their metrics match with those associated with COVID-19 or other diseases~\cite{seshadri2020wearable}. The wearable devices can also be used for broadcasting health knowledge~\cite{waheed2020successful},  providing capabilities for tracking and contact tracing~\cite{lin2020combat}, ensuring social-distancing~\cite{waheed2020successful}, and providing mental healthcare~\cite{ueafuea2020potential} by tracking an individual's cognition and mood in real-time, thus, enabling personalised interventions~\cite{sahakian2020covid}. These and other similar applications are boosting the demand of wearables. 
\citet{papa2020health} provide insights as to how the smart wearables would bring a revolution in healthcare. 
Research by Global Data~\cite{GlobalData} estimates that the market for wearables is now on track to increase from being worth nearly \$27 billion  in 2019 to a whopping \$64 billion by 2024. Next, we briefly discuss some recent relevant success stories in the fight against COVID-19.

WHOOP Inc. in collaboration with some leading research organizations has developed a COVID-19 identification system based on their WHOOP strap which measures respiratory rate using Resting Heart Rate (RHR) and Heart Rate Variability (HRV). The data from WHOOP strap is communicated to a mobile application and then forwarded to a cloud, collectively called WHOOP system \cite{IEEEhowto:Whoop}. Their system identified 20\% of COVID-19 positive individuals in the two days prior to symptom onset, and 80\% of COVID-19 positive cases by the third day of symptoms. 
\change{Philips has also developed disposable patches for early detection of COVID-19 patients~\cite{PhilipsPatches} and disposable biosensors for early COVID-19 patient deterioration detection~\cite{PhilipsBiosensors} which measure and transmit various predictors of deterioration such as respiratory rate, heart rate, activity level, posture and ambulation etc.}
 
Researchers from MIT and Harvard University investigated whether
COVID-19 subjects  could be accurately
discriminated only from a forced-cough cell phone recording using
AI~\cite{laguarta2020covid}. Their  results are based on cough recordings of over 5000 subjects and demonstrate that their  model  discriminates officially tested COVID-19 subjects
97.1\% accurately,  with a 100\% asymptomatic detection rate. In the past, cough recordings had also been used to accurately diagnose conditions such as pneumonia and asthma~\cite{porter2019prospective}. This shows the potential of integrating such solutions in wearables to enable a non-invasive, real-time solution for disease diagnosis, pre-screening, and outbreak monitoring.

\sloppy
The Scripps Research Translational Institute has conducted a study named DETECT (Digital Engagement \& Tracking for Early Control \& Treatment) that
collects data from smartwatches and activity trackers of consenting partners as well as self-reported symptoms and test results~\cite{IEEEhowto:Wearable}. They recently reported~\cite{quer2020wearable} that 
the data from wearable devices along with self-reported symptoms can be used for identifying cases of COVID-19  with greater success than looking at symptoms alone. 
To help combat COVID-19, many other  similar studies~\cite{mishra2020early,natarajan2020assessment,Evidation,TemPredict,RKI} have accelerated deployment to allow interested individuals to voluntarily share their sensor and clinical data.

\subsubsection{Changes in Regulations and Procedures}
In the past, adoption of IoT in healthcare had been slower than expected due to regulatory policies related to privacy, data security and approval procedures. In response to COVID-19, emergency regulations are being adopted using fast-tracked procedures and many new technologies have been given emergency approvals, which is accelerating IoT adoption. 
For example, U.S. Food and Drug Administration (FDA) has issued an Emergency Use Authorization certificate to the electrocardiogram low ejection fraction tool developed by Eko, a digital health company, to help clinicians assess cardiac complications associated with COVID-19~\cite{EKO}. Handheld and portable ultrasound solutions in particular have become valuable tools for clinicians treating COVID-19 patients due to their imaging capabilities, portability and ease of disinfection.
Philips has received clearance from FDA for Lumify, a portable ultrasound device~\cite{Lumify}. 
Lumify has a transducer which needs to be connected with  the user's smartphone running Lumify application. Moreover, FDA has also allowed Aidoc (a technology company) to use their AI-CT algorithms for COVID-19 detection~\cite{Aidoc}.

The World Health Organisation (WHO) has warned of the risk of infections in crowded hospitals and emergency rooms~\cite{barr2020telemedicine}. Consequently, regulatory response to minimise hospital and clinic visits has accelerated the adoption of telehealth and home care. For example, many countries are encouraging the use of telehealth services and have added various types of medical services on their public health programs (e.g., Medicare Australia) that can now be accessed via telehealth~\cite{AusMedicare_Telehealth}. The sharp increase in  the use of telehealth is encouraging the adoption of IoT and related technologies in healthcare sector. Analysis on the relationship between IoT technologies, smart telemedicine diagnosis systems, and virtual care for different age, race/ethnicity, gender, education, and geographic regions is presented in~\cite{clarke2020remote}. Similar analysis is performed in~\cite{hughes2020artificial, davis2020integrating} for AI, Internet of Medical Things (IoMT) and Big Data enabled healthcare for monitoring, detection and prevention of COVID-19. Recently, a novel IoMT platform has been proposed for remote health monitoring offering emotional treatment suggestion to the doctors and patients~\cite{zhang2020emotion}. A detailed analysis of contactless health services in pre-, during-, and post-pandemic periods can be found in~\cite{lee2021opportunities}. The study predicts that hybrid healthcare services would emerge in the post-COVID-19 era, potentially with new advances due to the accelerating technological development.

There are various other changes in regulations and procedures that are accelerating the adoption of IoT.
For example, to combat COVID-19, 
some countries have introduced confinement measures and tracking of COVID-19 patients using GPS, bracelets or other technological tools~\cite{OECD}. Furthermore, many countries have introduced new regulations including stricter cleaning requirements~\cite{cleaning_regulations} and record keeping requirements for contact tracing~\cite{tracking_regulation}. This has led to the deployment of robots for efficient and effective disinfection~\cite{RobotsCovid,guettari2020uvc,ackerman2020autonomous,zeng2020high} as well as technology-assisted record keeping and visitor tracking/monitoring~\cite{ahmed2020survey,xia2020return}. 
We provide more details of these in \autoref{sec:building} (Smart Buildings) because these IoT solutions (disinfectant robots and visitor tracking) are mainly deployed within buildings.

\end{paracol}
\nointerlineskip
  \begin{figure}[!t]
\centering
\widefigure
\includegraphics[width=\textwidth]{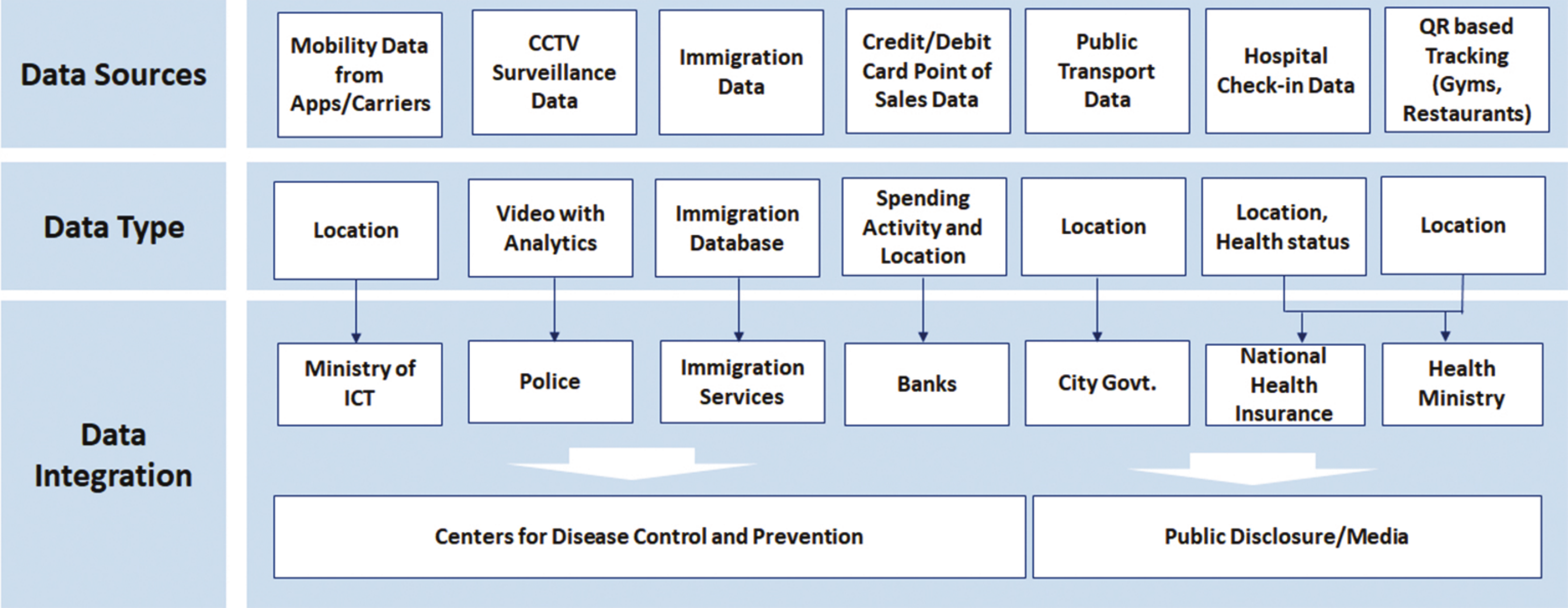}
\caption{An Example of a Test and Trace IoT Architecture}
\label{fig_taiwan}
\end{figure}
\begin{paracol}{2}
\switchcolumn
 
\subsubsection{IoT for Test and Trace}
IoT-enabled testing and tracing of COVID-19 can reduce the transmission which is critical in the fight against this pandemic. Therefore,
IoT is now being extensively used for test and trace applications which is accelerating IoT adoption~\cite{ContactTracingIoT}. For example, see
\cite{ribeiro2021towards} which provides an analysis of 13 technologies, to monitor users with COVID-19 symptoms, that can be used by governments and other organisations to combat COVID-19 and future pandemics.
A large number of countries and regions have adopted digital test and trace efforts, including UK, South Korea, Germany, Spain, Vietnam, Taiwan, etc.~\cite{moon2020optimising,cao2020localized}.  While some of these efforts have not been very successful in tracing COVID-19, the countries that have extensively used technology in their solutions have done much better in their fight against this pandemic. \change{For example, it was reported in \cite{o2020covid} that UK's test and trace system employing around 27,000 contact tracers was not able to reach 21\% of the people who were tested positive in the week of 2 July to 8 July 2020. Moreover, the 79\% of the people who were reached identified 13,807 close contacts but only 71\% of these contacts were reached and asked to self-isolate. On the other hand,  
Taiwan had better systems in place (due to the lessons learned from the 2003 SARS outbreak) and was successful in curbing the first COVID-19 wave mainly due to quick and effective policy decisions and extensively using digital technologies~\cite{IEEEhowto:Policy_Taiwan}. Despite Taiwan's initial success, there has been a surge of COVID-19 cases in Taiwan recently which is attributed to the lax adherence to safety protocols, fewer tests (around 13 total tests performed per thousand people as of 21 May 2021, in contrast, US has conducted around 1322 tests per thousand people~\cite{OurWorldInData}) and slow vaccine roll out (only 1.26\% of the population has received a COVID-19 vaccine as of 21 May 2021~\cite{OurWorldInData}). This shows that an advanced technology-based test and trace system can only be an effective tool to fight against the pandemic if all other safety protocols are also followed and strict policy decisions are enforced.} 

\autoref{fig_taiwan}
shows an architecture diagram for a test and trace system similar to the solution employed by Taiwan.  There are mainly two layers of such advanced test and trace system, namely Data Acquisition and Data Integration.

\noindent\textbf{Data Acquisition.} Mobility data is acquired from multiple sources including immigration department database, GSM and GPS based mobile phone tracking, and QR code based tracking etc. This helps Taiwan in tracking international as well as local travellers. Resources from smart city are also utilized including CCTV cameras based surveillance~\cite{shorfuzzaman2020towards}. Credit card based payments are also detected and tracked to locate users and predict their activities~\cite{CreditCardTracing}. Similarly, hospital visits of the individuals are also recorded and tracked to help contact tracing~\cite{IEEEhowto:Response_Taiwan}.

\noindent\textbf{Data Integration.} In this layer, the data from different sources are integrated and shared with the relevant departments \cite{IEEEhowto:Taiwan_actions}. For example, the immigration database and  National Health Insurance (NHI) database are shared with the Centers for Disease Control and Prevention~\cite{IEEEhowto:Policy_Taiwan}. Location information is shared with local authorities. For example, surveillance data from smart city setup is shared with police officers for necessary actions~\cite{IEEEhowto:Policy_Taiwan}. Data extracted from certain sources such as public transport and shopping malls is also shared with public to help them make informed decisions about their daily routines~\cite{transitAuthority}.

\subsection{Smart Homes}

ABI Research predicts that, despite the economic disruption by COVID-19, smart homes revenue will reach \$85 billion in 2020, an increase of 4\% over 2019~\cite{ABIResearch_SmartHomes}. However, this is well below the pre-pandemic prediction that forecasted a growth of 20\% in 2020. Fortunately, this negative impact is predicted to be temporary and the pandemic is expected to have a positive impact in the long-term with revenue reaching \$317 billion by 2026, an increase of 5\% over pre-pandemic forecasts~\cite{ABIResearch_SmartHomes}. This is mainly because the changes in lifestyles and consumer behaviors that will drive adoption are here to stay even after the pandemic is over. For example, many organisations have announced that they will be adopting work from home as a long-term policy. Microsoft, Twitter, and Square have announced that workers can continue working from home even after we overcome the pandemic. A Gartner survey points out that 74\% of CFOs are already considering work from home as a permanent policy because of cost savings~\cite{GartnerSurvey}. 
A survey from Redfin shows that more than 50 percent of New York, Seattle, San Francisco and Boston workers would move, in order to save on rent and cost of living, if work from home became a permanent option~\cite{RedfinSurvey}.

As more people will be working/studying from home, they are expected to adopt various IoT devices to create a more comfortable work/study environment.
For example, in Australia, smart speakers, security lighting, and energy \& HVAC sensors are showing strong sales growth in 2020~\cite{Telsyte}.
\autoref{fig_smarthome} summarizes some current and expected developments in smart homes as a consequence of COVID-19.
Below, we give the details of some of these.

\begin{figure}[!t]
\centering
\includegraphics[width=0.9\columnwidth]{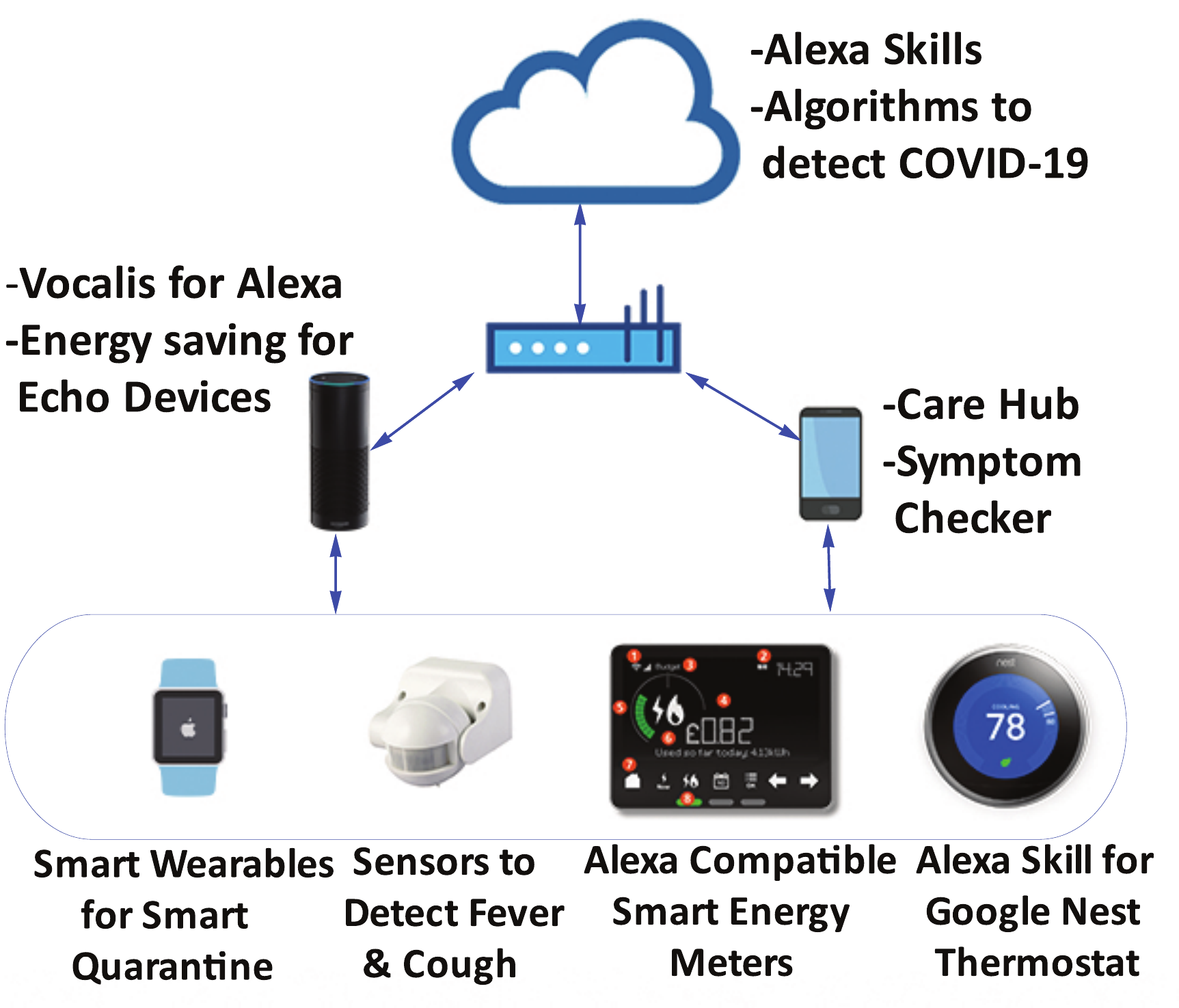}
\caption{Recent Developments in Smart Homes in the Wake of COVID-19}
\label{fig_smarthome}
\end{figure}

\subsubsection{New Features and New Sensors}
Many Smart Home devices have introduced various new features in response to COVID-19. For example, Amazon's Alexa Care Hub can help people to remotely check in on their family members who they are not able to visit due to COVID-19 restrictions~\cite{CareHub}. When family members agree to connect on Care Hub, their activity profiles on Echo devices are shared with each other and alerts/notifications can be generated for certain situations. Infermedica, a health tech startup,  has developed a mobile application called Symptom Checker which interviews home occupants and guides them about their current health. They have also developed an Alexa skill to integrate Symptom Checker with smart homes~\cite{SymptomChecker}. Technologies and tools are also being developed to diagnose COVID-19 using speech analysis. Vocalis, a technology company providing voice-based healthcare solutions, is developing a mobile application (Alexa skill will be developed later) which will be able to detect symptoms of COVID-19 using voice signals~\cite{Vocalis}. They had previously developed solutions to detect flare-ups of chronic obstructive pulmonary disease by identifying the signs that users were short of breath when speaking. WiFi signals have also been used as a non-intrusive approach to measure respiratory rate of COVID-19 patients~\cite{li2020wi}.
Alexa is now also connected with wearables~\cite{AlexaWearables}. These new features are helping researchers to propose smart quarantine solutions ~\cite{alrumayh2020supporting}. \change{\citet{nguyen2020comprehensive} provide a comprehensive overview of how different technologies and sensors (e.g., machine learning, computer vision, thermal sensors, inertial sensors etc.) can be used to encourage or enforce social distancing. They suggest that smart infrastructures should incorporate a \emph{pandemic mode} in their design to better respond to future pandemics.} 
IoT systems to detect fever, cough, sneezing, etc. are also being proposed to better analyze the health profile of home occupants~\cite{arun2020detection}. These and other similar advancements enable web dashboards and mobile applications to better monitor and analyse the health profiles of smart home residents.  

\subsubsection{New Home Energy Management Solutions}
Homes are increasingly being equipped with smart energy management systems. Nowadays, homes typically have multiple power sources such as solar, battery bank, and grid~\cite{HybridEnergy}. Due to COVID-19, daily routines of people have changed and they spend more time at home. So, domestic energy consumption has increased significantly \cite{HEMS}. 
To help save energy, Amazon has introduced new features in Alexa to keep users updated about the energy consumed by Echo devices and suggest energy saving features~\cite{AlexaEnergy}. COVID-19 is acting as a trigger for smart meter installations in many countries. Energy Efficiency Services Limited (EESL), the agency responsible for the implementation of smart meters in India, reports that smart meter adoption in India has increased since the COVID-19 outbreak~\cite{SmartMetersIndia}. Some energy suppliers have also started offering Alexa-based services for managing energy utilities at home~\cite{EDF}. Similarly, smart thermostats are seeing rapid expansions~\cite{SmartThermostat}, e.g., Google Nest Thermostat has introduced new feature named Savings Finder to save energy at homes~\cite{GoogleNestThermostat}. 

\subsubsection{New Features to Assist xFH}
Home confinement due to COVID-19 has led researchers and developers to explore new ways and features to assist different modes of ``x from home'' (xFH), i.e., work from home, education from home, shopping from home, and healthcare from home etc. \citet{lm2020impact} have enlisted impacts of COVID-19 on education and tools available for online-cum-distance learning. 
Similarly, to assist work from home, Google has launched Google Workspace which is a rebranding of existing products with new features~\cite{GoogleWorkspace}. It supports creating new documents within a room in Chat, Google’s Slack-like chatrooms, without having to switch tabs. It also allows videoconferencing while collaborating on documents, sheets, and slides. People are also investing time and money in setting up home offices by using IoT-related technologies (e.g., smart lights, smart thermostats) to create smarter and more comfortable work environments. Researchers are also proposing new healthcare systems that enable remote monitoring of patients' health from their homes~\cite{ShHeS}.

\subsection{Smart Buildings}\label{sec:building}

COVID-19 has had profound impact on built environments~\cite{megahed2020antivirus}. Strict lockdown measures have significantly reduced foot traffic in buildings which has resulted in bankruptcies of various notable offline stores including Ascena Retail Group and Brooks Brothers. Malls are being converted to warehouses mainly because of the growth of e-commerce in the wake of pandemic~\cite{galhotra2020impact}. One of the largest owners of shopping mall real estate in the USA, Simon Property Group, has been talking to Amazon about transforming its anchor department stores into Amazon distribution hubs~\cite{AmazonSimon}. Microsoft Azure suffered capacity shortages of data centers in Europe. So, more data centers are being built to meet the increased demand of the ``x from home'' trends~\cite{MoreDataCentres}.

COVID-19 has also fundamentally changed the way buildings are expected/required to operate, e.g.,  social distancing, occupancy tracking/monitoring, smart Heating, Ventilation and Air-Conditioning (HVAC) systems and stricter cleaning requirements. 
This has significantly increased the importance and demand of IoT in buildings~\cite{ASMAG_imppact_of_Covid,CMS_smart_building} because smart buildings can enable more efficient facilities management and help support a safe and healthy environment~\cite{DeloitteSmartBuilding}. However, despite the increased importance and demand, investment in smart buildings has been mostly negatively affected in the short-term mainly due to the financial stress on organisations. Consequently, it is predicted that market for global smart buildings is expected to decline from \$49.23 billion in 2019  to \$41.72 billion in 2020. However, this is expected to recover and reach \$54.94 billion in 2023~\cite{ReportLinker}. 

We remark that while the overall investment in smart buildings is expected to decrease in the short-term, certain applications of smart buildings (e.g., safety and health applications) are expected to see increased investment even in the short-term. A recent global corporate survey of 250 executives conducted by Verdantix~\cite{Verdantix} found that, while the COVID-19 outbreak has caused 24\% of organizations to slow down their digitization, 47\% are planning to make new investments in workplace safety technology, including investment in thermal imaging technology, mobile workplace apps and space monitoring analytics. 
\autoref{fig_smartbuilding} shows a summary of new initiatives across some major themes  significantly impacted by COVID-19 that are accelerating IoT adoption in smart buildings. We discuss the details below.

\begin{figure}[!t]
\centering
\includegraphics[width=\columnwidth]{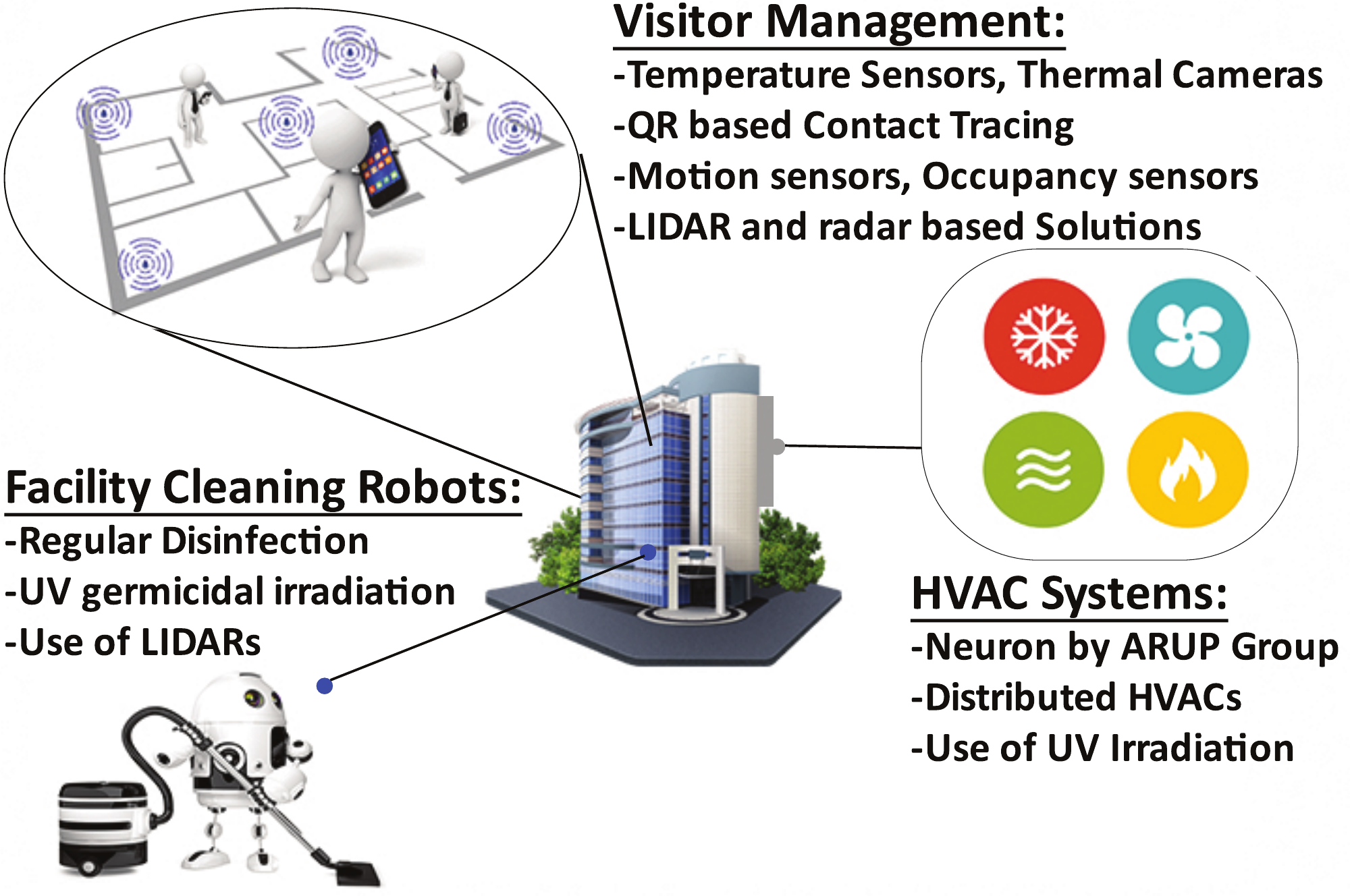}
\caption{Recent Developments in Smart Buildings in the Wake of COVID-19}
\label{fig_smartbuilding}
\end{figure}

\subsubsection{Smart HVAC Systems} 

Several studies have analysed the effect of ventilation, indoor air quality, lighting and the deposition on the surfaces of materials on the transmission of COVID-19~\cite{dietz20202019}. It is suggested that indoor air quality and HVAC systems can significantly reduce the airborne transmission of viruses~\cite{smieszek2019assessing,pinheiro2020covid}.
Consequently, best practices on HVAC to minimize the airborne transmission of the coronavirus in indoor environments have been proposed~\cite{HVAC}.
This is leading to a renewed focus on  smart HVAC systems in buildings~\cite{ArupSmartBuildings}.
HVAC systems are adopting UV irradiation for killing viruses while increased rate of air change is recommended to keep indoor air fresh~\cite{HVAC_UV}. \change{It has been shown that night ventilation can also help reducing the energy used for cooling buildings~\cite{guo2020optimization}.  Thus, a smart HVAC system may use an optimised ventilation strategy to improve indoor air quality while minimizing energy usage.}

Indoor air quality sensors can capture  level of pollutants, CO$_2$ and volatile organic compounds in indoor environment. Thus, IoT-enabled HVAC systems can be used to analyse air quality and formulate an action plan, track progress and assess mitigation effectiveness~\cite{ArupSmartBuildings}. For example, Arup Group developed a digital platform called Neuron~\cite{Arup_Neuron} that measures indoor air quality to predict or monitor high-risk conditions and allows remedies such as ventilation, UV light, or air purification to improve the indoor air quality. It can also
sense the body temperature to identify individuals with elevated temperatures and send them to separate queues for further screening.

\subsubsection{Visitor Management}

Health and safety of visitors are among the highest priorities for the businesses and organisations that are open/reopening. In fact, many states now legally require organisations to implement COVID safe plans to prevent the spread of COVID-19~\cite{VIC_COVID_SAFE,NSW_COVID_SAFE}. This has increased the demand of smart visitor management to enable social distancing, contact tracing, and occupancy tracking/monitoring. For example, office buildings are installing temperature sensors and thermal cameras to detect employees with fever~\cite{thermalCamera}. QR code based contact tracing is being used in public places such as gyms, restaurants and coffee shops~\cite{QR}. Proximity sensors based fast contact tracing has been proposed in~\cite{xia2020return}. A review on the recently developed contact tracing mobile applications has been provided in~\cite{ahmed2020survey}. Techniques are being proposed to visualise distance between visitors in indoor environments~\cite{socialDistance} and to provide indoor navigation for visitors while ensuring social distancing~\cite{fazio2020proximity}. Motion sensors, occupancy sensors and distance sensors are also being installed to track real-time status of office occupancy. Proofs of concept are being implemented by using LIDAR and radar to ensure social distancing without compromising the privacy of visitors~\cite{fan2020autonomous, sathyamoorthy2020covid}.

\subsubsection{Facility Management and Cleaning} 

COVID-19 demands improved facility management and cleaning capabilities to avoid the virus spread. New SOPs are being adopted for cleaning and disinfection~\cite{Practices_Cleaning}. Smart buildings should include efficient and effective disinfection of indoor spaces especially frequently used surfaces such as door handles, and elevator buttons and switches. To this end, companies are deploying robots to disinfect venues frequently and efficiently~\cite{RobotsDisinfect}.
Researchers at the Computer Science and Artificial Intelligence Laboratory (CSAIL) at MIT have developed a robot to ensure spaces are kept as clean as possible~\cite{CSAIL_MIT}. The robot works on the principle of ultraviolet germicidal irradiation and utilizes short wavelength ultraviolet light to kill microorganisms. Similarly, UVD Robots, a subsidiary of Blue Ocean Robotics, have developed a robot equipped with mobile base and LIDAR~\cite{ackerman2020autonomous} that creates a navigation map of patient rooms to be disinfected. Furthermore, the data from various sensors deployed in a smart building can be used to identify the spots that need to be disinfected, e.g., areas visited by many people or surfaces touched by someone who is detected to have elevated temperature or other symptoms. Robots developed by Boston Dynamics~\cite{RobotsCovid} are being used in Brigham and Women’s Hospital of Harvard University. Researchers in~\cite{guettari2020uvc} have developed an i-Robot  equipped with eight UVC lamps for disinfection purposes. A recent study~\cite{zeng2020high} elaborates on disinfection robots being used in different types of buildings.

 \subsection{Smart Cities}

COVID-19 has affected many facets of urban life including reduced traffic and public transport usage, changes in energy demand patterns, disruption in supply chains and increase in web traffic etc.
According to a recent report from ABI Research, governments are striving to improve their urban resilience and  digital transformation to fight against COVID-19 and its impacts~\cite{ABIResearch_SmartCities}. Drones, new types of surveillance, digital twins, real-time dashboards and autonomous freights are among the technologies being deployed for the new use cases emerged during the pandemic~\cite{ABIResearch_SmartCities}. Despite the financial strain, governments are investing heavily in technology to help combat COVID-19~\cite{ARCAdivsoryGroup}. For example, to boost the economy and support businesses recovering from COVID-19, the Government of Singapore announced an increase in its investment towards digitalisation by 30 percent increasing from \$2.7 billion in 2019 to \$3.5 billion in 2020~\cite{Singapore_IncreasedInvestment}. China has also announced a dramatic increase in investments in IoT and related technologies such as 5G, smart grids, and data centers.  
\autoref{fig_smartcity} summarizes the impact of COVID-19 on smart cities. Next, we discuss the details of some of the key initiatives and factors driving IoT adoption in smart cities.

\begin{figure}[!t]
\centering
\includegraphics[width=\columnwidth]{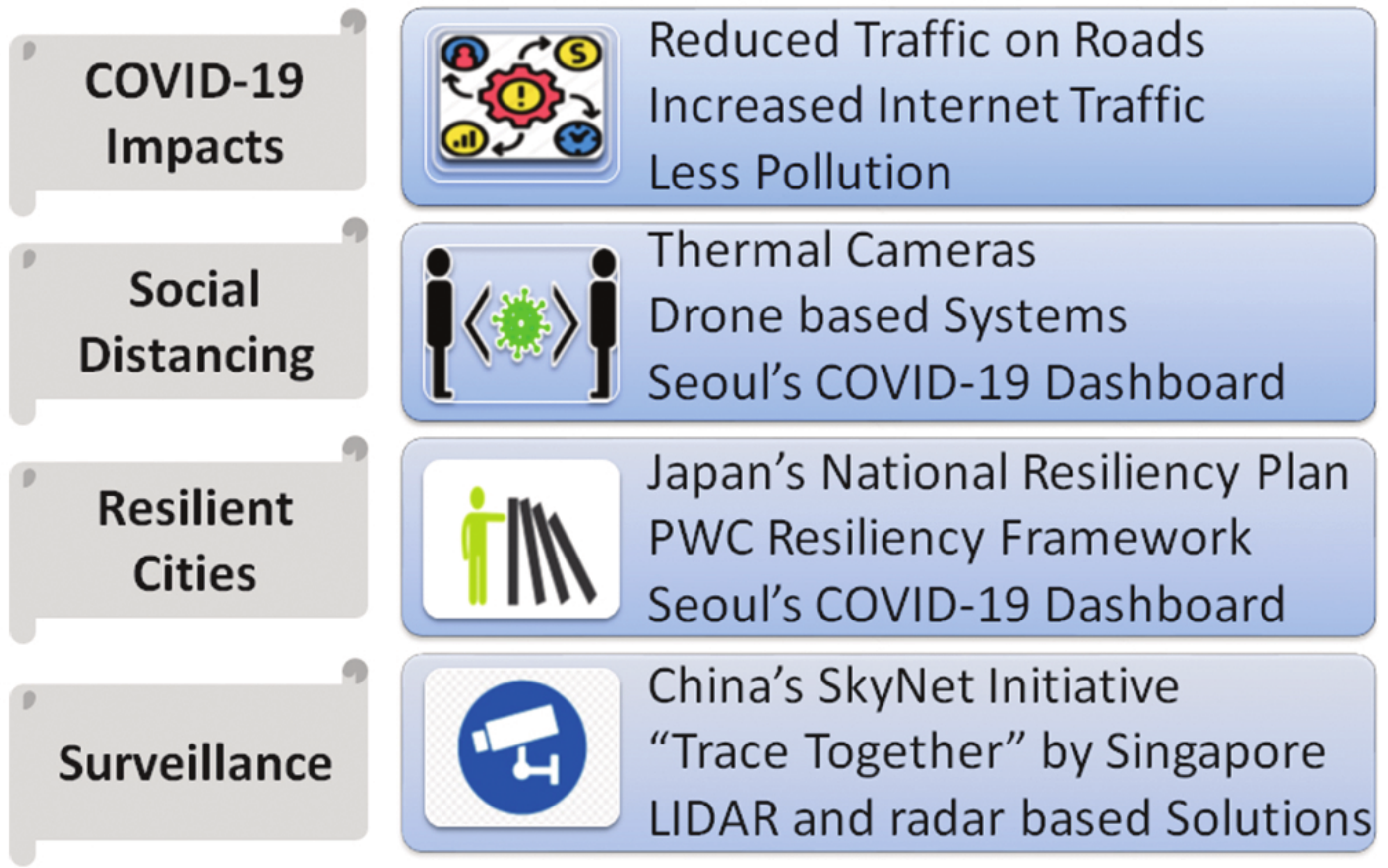}
\caption{Recent Developments in Smart Cities in the Wake of COVID-19}
\label{fig_smartcity}
\end{figure}

\subsubsection{Initiatives From Some Leading Smart Cities}
By definition, cities that are ``smarter'' are more likely to be better prepared for disasters and pandemics, e.g., they may already have an established crisis operation centre connected to a number of IoT devices such as sensors and cameras~\cite{MSCI_SmartCity}. 
Some of the leading smart cities (e.g., Hong Kong, Singapore, Seoul, Taipei) had already invested heavily in IoT and related technologies before the pandemic hit. For example, in 2017, 
Hong Kong’s Smart City Blueprint outlined its mission to make the city more livable, sustainable, resilient and competitive.  
Consequently, these smart cities were able to act more swiftly and effectively during the COVID-19 crisis~\cite{MSCI_reflections}.
Below, we briefly discuss some initiatives by some of the leading smart cities in their fight against COVID-19.

Singapore has launched a mobile application called TraceTogether that helps with contact tracing by tracking the events when people with the application installed in their mobile devices are in close vicinity of one another~\cite{Singapore}. The mobile application is also capable of scanning ``Safe Entry'' QR code at the entry point of public places. The data generated is helpful in tracking the coronavirus in the city. They have also introduced contact tracing using wearable technology and are planning to distribute wearables nationwide for contact tracing. Helsinki is using Laboratory registry data for the categorized surveillance of COVID-19 suspects~\cite{Helsinki}. Seoul has developed a dynamic COVID-19 dashboard containing important information of the confirmed COVID-19 patients including their mobility details. This information is updated multiple times throughout a day and is helpful for the individuals to take precautionary measures~\cite{Seoul}. 
Taiwan's ability to quickly implement additional arrival screening as well as enforcing contact tracing has helped keep mortality rate low~\cite{MSCI_reflections}.

\subsubsection{Enhanced Sensing Coverage}
More sensors are being installed in smart cities since the beginning of COVID-19 pandemic. China has capitalized well on its SkyNet initiative. More than 20 million cameras installed under this project have been used in identifying and monitoring densely populated areas in cities \cite{Skynet}. Some countries have adopted more privacy-aware monitoring solutions based on radars and LIDARs~\cite{VelodyneLidars}. Thermal cameras for fever detection have seen 83 percent growth in 2020~\cite{IEEEhowto:FeverDetection}. Drones are not only being used for contactless delivery~\cite{DroneStartups} but are also being used for monitoring social distancing~\cite{Dronebased} in many countries such as Israel, UK and Singapore.
Apart from enhanced sensing coverage, an advanced surveillance system to fight against COVID-19 is also the need of the hour. Many researchers have proposed new cloud-based solutions for effective surveillance of probable cases of COVID-19. The Honghu Hybrid System (HHS)~\cite{HHS} is designed for the collection, integration, standardization, and analysis of COVID-19 related data from multiple sources. HHS includes a case reporting system, diagnostic labs, electronic medical records, and social media on mobile devices. Such systems can also be used for data analytics for regulating foot traffic and to trigger warnings for crowds.
  
\subsubsection{Resilience in Smart Cities}
COVID-19 is not the only catastrophe which we have ever faced or will face. There are always possibilities of future pandemics, recurring natural disasters (e.g., earthquakes, floods and droughts) and disruptions caused by humans (e.g., strikes, cyber attacks and terrorism). It should not be very surprising that urban areas are typically more vulnerable to disasters, e.g., around 95\% of the COVID-19 cases have been reported in urban areas. 
COVID-19 has proven to be a catalyst to ensure resilience in smart cities and there has been an increased interest in developing smart cities that are economically and environmentally resilient against natural disasters and human created troubles.  \change{PwC Middle East has derived a smart city resiliency cyclic framework consisting of Sense (predict threats using data analytics and other tools), Defend (reinforce weak links within city organisations), Respond (set up crisis management and empowered task forces), and Recover (identify key assets that need to continue operating at a reduced capacity)~\cite{PWC_resiliency}. PwC Middle East selected five cities -- Boston, Helsinki, Riyadh, Singapore and Vienna -- that adopted smart technologies and demonstrate how the proposed framework can be applied in this pandemic and to build long-term resilience against future threats. 
} Japan has also set an example in implementing smart cities resiliency. Its National Resiliency plan covers smart communications,
sustainable energy systems and resilient water networks. Initiatives taken in Japan for National Resiliency has been enlisted in~\cite{dewit2020integrated}.
 \subsection{Transportation}

Smart transportation market was growing at a substantial pace before COVID-19 and was projected to grow from \$94.5 billion in 2020 to \$156.5 billion by 2025~\cite{TransportationMarket}. COVID-19 is a threat to this significant expected growth and has already slowed down the rise of this market. 
Both passenger transport and freight have suffered severe setbacks from the COVID-19 pandemic~\cite{tardivo2021covid}.
A steep decrease in number of trips on public transport has been seen in the last year or so. Subways are also seeing a sharp decline in number of passengers. Moreover, delays in shipments have also been seen due to closure of ports in many countries. Road transport remained operational but with minimal itineraries. Theses changes have resulted in increased volatility of freight charges, according to European Road Freight Rate Benchmark Q1 2020 report~\cite{RoadMarket}.

\begin{figure}[!t]
\centering
\includegraphics[width=0.8\columnwidth]{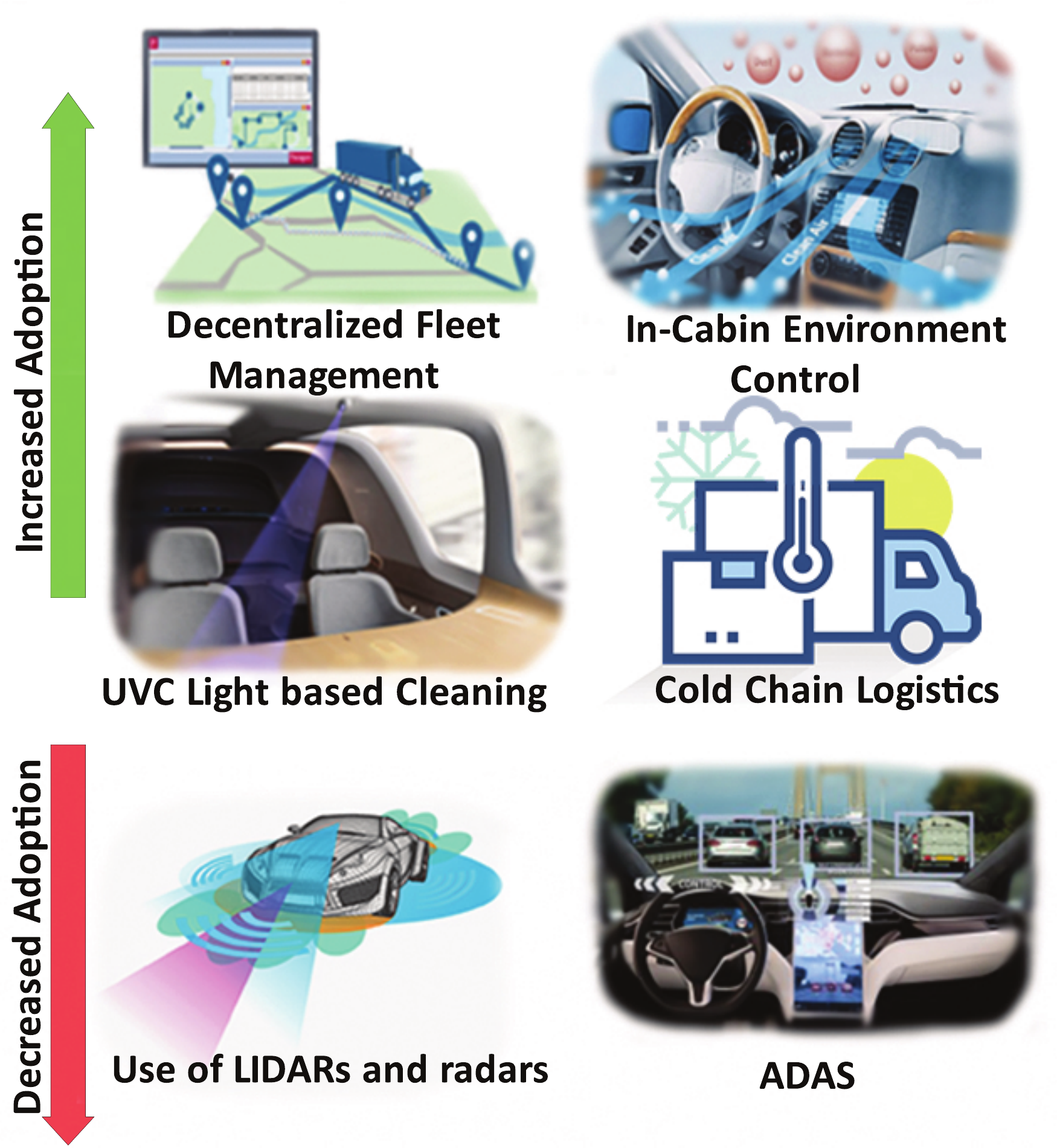}
\caption{Summary of IoT Adoption in Transportation}
\label{fig_automotive}
\end{figure}

Automotive industry was already in recession even before COVID-19. The pandemic has created another dip to this recession which is expected to delay the recovery until 2025~\cite{RecessionAutomotive}.
The current recession has resulted in a delay of Autonomous Driving (AD) and Advanced Driver Assistance Systems (ADAS) adoption in cars \cite{DelayADAS}. Research and development projects are being scaled down and corresponding timelines have been adjusted. Most LIDAR and radar startups in ADAS space are pivoting to non-automotive use cases such as industrial automation and mobile devices. Specifically, introduction of LIDARs in iPhones by Apple has re-ignited the interest of LIDAR players in smartphone and consumer space where they expect higher volumes in shorter time because of shorter design cycles in consumer segments~\cite{iphoneLidars}. 
Although the transportation sector has been hit pretty hard by the pandemic, there are still some areas within the sector that are attracting increasing attention such as in-cabin analytics and smart logistics.
\autoref{fig_automotive} shows the summary of these. Below, we give the details.
\subsubsection{In-Cabin Sensing and Analytics}
Modern cars are increasingly being equipped with climate control features including in-cabin air quality control systems. Sensors and technologies are being deployed in vehicles to measure CO, CO\textsubscript{2} and other pollutants within the vehicles~\cite{IncabinControl}. It was reported that low doses of far-UVC light inactivate airborne coronaviruses without harming human tissues~\cite{manuela2020far}. Several automakers including GM and Hyundai have proposed the use of UV light to clean the vehicle cabins~\cite{InCabinUV}. Uber has introduced facial mask detection in their navigation application to assure riders that the drivers are following SOPs~\cite{Uber}.
Techniques are also being developed to monitor indoor air quality and occupancy in public transport, which is crucial to curb the transmission of COVID-19~\cite{nurettincovid}. Many jurisdictions are deploying IoT solutions to monitor real-time occupancy on public transport. This data is also often shared with passengers to help them make informed decisions and comply with social distancing requirements~\cite{PTV_VIC_IoT}.

\subsubsection{Smart Logistics} 
Smart logistics market is expected to grow at a compound annual growth rate (CAGR) of 8.5\% from 2020 to 2027~\cite{LogisticsMarket}. This is partly due to the dire need of IoT in smart logistics created by COVID-19. China is the world's largest logistics market and it has also responded well to COVID shock. Cainiao Logistics has built China Smart Logistic Network based on IoT and AI to achieve fast express delivery services~\cite{liu2020china}. China's smart logistics developments in wake of COVID-19 have been detailed in~\cite{wang2020analysis}. XPO Logistics, a world's leading logistics company, has enhanced its existing smart logistics platform named XPO Connect by introducing various new features including enhanced inventory tracking precision and contactless delivery~\cite{XPO}. Since the beginning of COVID-19 pandemic, cold chain logistics have been emphasized by many scientists and researchers. It will not only allow safe transport of coronavirus vaccines but also help existing supply chain systems to better preserve food and pharmaceutical items during transportation. \citet{belhadi2021manufacturing} categorize existing research on supply chain resilience into two main categories: proactive strategy and reactive strategy. Further, it proposes a framework to explore resilient supply chain strategies that include evaluation of the impact that COVID-19 has caused on the global supply chains. Researchers are working on enhanced IoT adoption in cold chain logistics~\cite{masudin2020food, tsang2020integrating, wu2020experimental}.
Smart fleet management solutions are also gaining more attention. IoT enabled connected fleets reduces the operational hours and allows decentralized operations~\cite{DecentralizedFleet}.
Researcher are also proposing optimization procedures to operate minimum number of fleets for maximum utility~\cite{basso2020public}. \change{Smart logistics solutions are also needed for effective vaccine roll-out. NearForm has come up with a proof-of-concept vaccination app to improve the management of inventory and resource allocation for COVID-19 vaccines~\cite{NearForm}. Vodafone has offered to support COVID-19 vaccine roll-out in Africa using cellular and IoT tracking and monitoring technologies~\cite{Vodafone-rollout}.}

\begin{figure}[!t]
\centering
\includegraphics[width=\columnwidth]{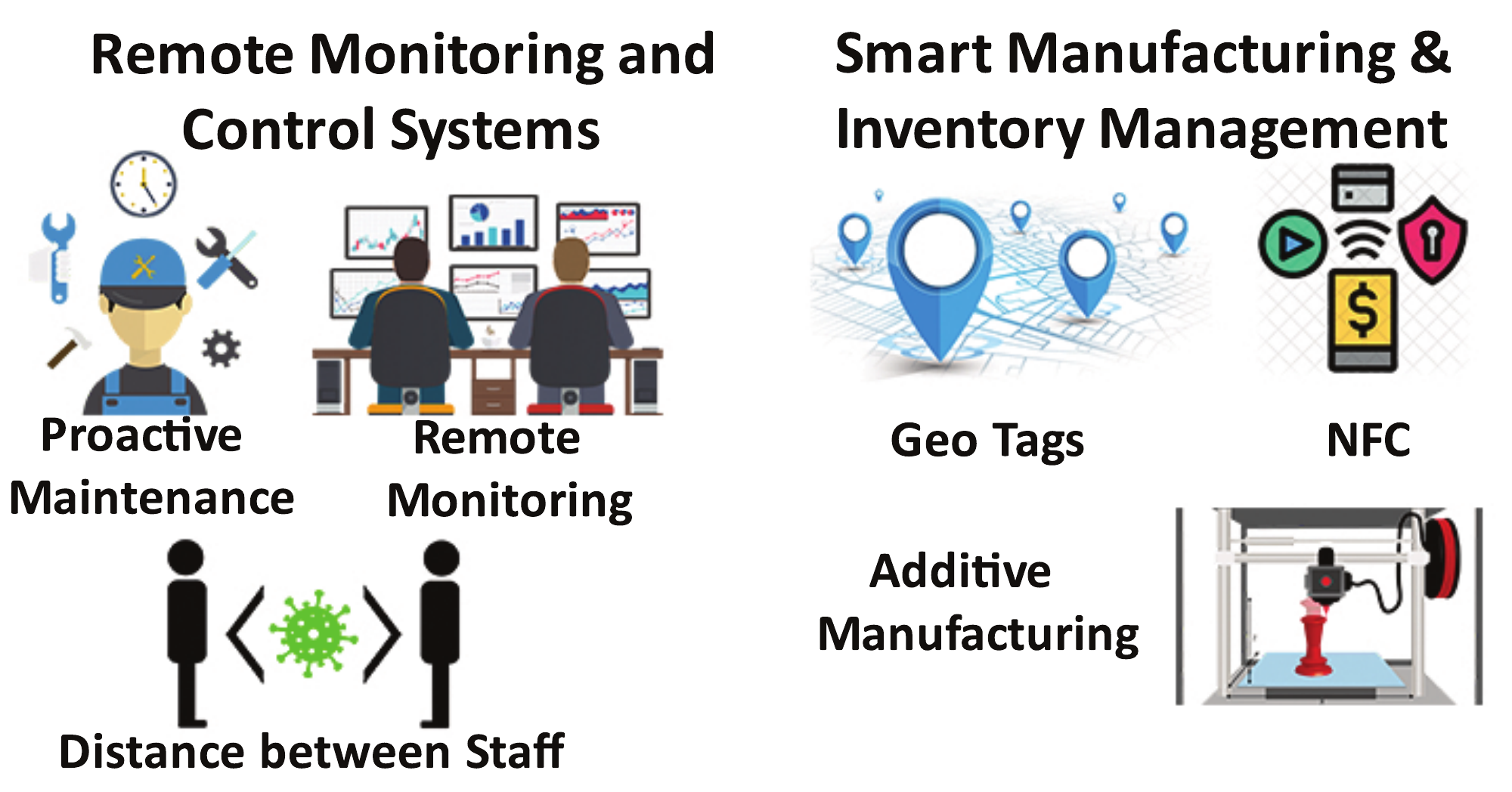}
\caption{Recent Increased Adoptions in Industries in Wake of COVID-19}
\label{fig_IndustrialIoT}
\end{figure}

\subsection{Industrial IoT}

Decrease in economic activity due to COVID-19 will have an adverse impact on factory automation as many companies are cutting on their industrial IoT (IIoT) budget. That being said, in mid- to long-term, IIoT is expected to see significant growth due to its importance especially during pandemics or other disasters.
A recent McKinsey survey found that 93\% of supply chain and manufacturing professionals plan to focus on smart and resilient solutions~\cite{IIoTResilientSupplyChains}. A report on India's initiatives for IIoT in wake of COVID-19 focuses on supply chain visibility and sustainable IIoT solutions~\cite{IIoTDeloitte}. It also shows how IIoT can have a bigger impact on returning industries back to normal as compared to other technologies like AI and Big Data. 
Thus, despite an adverse impact on IoT adoption in industries in the short-term, this pandemic is expected to result in  accelerated growth of IIoT in the long run. 
According to a research analysis, IIoT market will reach \$263.4 billion by 2027~\cite{IIoTMarket}. 
Similarly, Juniper Research forecasts that the number of IIoT sensors will increase from 17.7 billion in 2020 to 36.8 billion in 2025~\cite{IIoTJuniper}.  Next, we discuss recent developments in some key IIoT areas impacted by COVID-19. These are summarized in \autoref{fig_IndustrialIoT}.

\subsubsection{Remote Monitoring and Machine Health Prognostics}
IoT can help industries to continue production during pandemics. Examples of IoT solutions helping to fulfill SOP requirements (such as maintaining distance among factory workers and  intelligent equipment disinfection) can be found in recent literature~\cite{chen2020smart,zhang2020hybrid}. IoT is also used for remote monitoring and operation of the industrial machines~\cite{IEEEhowto:RemoteManufacturing}. 
The remote monitoring and operation need installation of various sensors in predictive maintenance space. One such system for textile sector is proposed in \cite{RMS}. Machine health prognostics has been a prime focus of many industries as lockdowns have led to loss of revenues. Remote identification and predication of machine failure have become even more important because on-site repair and maintenance are more expensive and challenging (or even impossible) during the pandemic. Recent efforts in developing a framework for machine health prognostics can be found in literature~\cite{atamuradov2020machine}. Different AI-based approaches to diagnose and predict a machine fault have also been proposed by many researchers. For details,  see \cite{MachineHealth} and references therein.

\subsubsection{Smart Manufacturing and Inventory Management}
Increased demand, fewer operational hours and limited number of on-site staff have made it difficult for the industries to  fulfill manufacturing demands. This is giving rise to additive manufacturing to meet the increased demand of various items such as personal protective equipment and other medical supplies~\cite{bragazzi2020digital}. The role of additive manufacturing in medical applications have been discussed in details in~\cite{patel2020role}. \citet{brem2021implications} detail how flexible manufacturing systems have been utilized by different countries to manufacture necessary medical equipment within a short duration of time. \citet{shen2020address} have proposed an intelligent collaborative framework to improve the resiliency and viability of the existing manufacturing processes. 

Due to the highly fluctuating demands, inventory management has become even more important.
A smart inventory management system that can monitor items and available space, and communicate it to the concerned warehouse team is much more needed during the pandemic. Geotags, ultrasound transducer and NFC are being used to build autonomous inventory management systems \cite{InventoryManagement}.

\end{paracol}
\nointerlineskip
\begin{specialtable}[htbp]
\caption{Summary of recent initiatives in the wake of COVID-19.\label{table:initiatives}}
\begin{tabular*}{\textwidth}{@{} llp{95mm} @{}}
\toprule
\textbf{Sectors} & \textbf{Categories} & \textbf{Initiatives} \\
\midrule
\textbf{Healthcare} & Diagnosis &- Whoop Strap to measure respiratory rate\newline- Philips' disposable patches to detect COVID-19
 \newline - Disposable biosensors by Philips
 \newline - DETECT by Scripps to collect data from wearables
 \newline - Taiwan's test and trace architecture
\\
&Regulations and Procedures & 
    - FDA approval for electrocardiogram low ejection fraction tool developed by Eko
    \newline - FDA approval for Lumify (a portable ultrasound device)
    \newline - Lenient regulations for telehealth services
    \newline - Stricter regulations for cleaning and disinfection procedures\\ \\
\textbf{Smart Homes} & New Features & 

     - Amazon's Carehub to connect with family members
    \newline - Alexa skill developed by Vocalis to diagnose COVID-19 using speech analysis
    \newline - Smart quarantine solutions
    \newline - Sensors to ensure social distancing
\\
&Energy Management Solution & 

    - Savings Finder by Google Nest to save energy
    \newline - Alexa services to manage energy utilities at home
\\ \\
\textbf{Smart Buildings} & Resilient Indoor Ambience & 
    - Neuron developed by Arup Group that measures indoor air quality to predict high-risk conditions
    \newline - UV irradiation adopted in HVAC systems
    \newline - Robot-based facility cleaning
\\
&Visitor Management & 
    - QR code based contact tracing
    \newline - Proximity sensors based fast contact tracing
    \newline - LIDAR and radar based solutions to ensure SOP measures
\\ \\
\textbf{Smart Cities} & New Initiatives & 

    - TraceTogether developed by Singapore
    \newline - Helsinki's labortary registry data used for categorized surveillance of COVID-19 suspects
    \newline - Seoul's dynamic COVID-19 dashboard
\\
&Enhanced Sensing Coverage & 
    - Thermal cameras for fever detection
    \newline - Drone based contactless delivery and surveillance. 
\\
&Resiliency & 
    - PwC's smart city resiliency cyclic framework
    \newline - Japan's national resiliency plan
\\ \\
\textbf{Transportation} & In-Cabin Features & 
- UV light based cleaning of vehicle cabins
\newline - Facial mask detection in Uber application
\newline - Occupancy monitoring in public transport \\
&Smart Logistics &- Smart logistic network developed by Cainiao
\newline - New features in XPO connect
\newline - Smart cold chain logistic solutions for vaccine
\newline - Smart fleet management during lockdown\\ \\
\textbf{Industrial IoT} & Remote Monitoring & - Machine's remote health prognostics\\
&Smart Manufacturing & 
- Enhanced role of additive manufacturing
\newline - New resilient manufacturing frameworks
\newline - Autonomous inventory management systems\\
\bottomrule
\end{tabular*}
\end{specialtable}

\begin{paracol}{2}
\switchcolumn

\change{\autoref{table:initiatives} summarizes this section and presents a summary of recent IoT initiatives taken in wake of COVID-19 in different sectors.}

\section{Challenges and Key  Research Directions}\label{sec:challenges}

COVID-19 has brought about opportunities as well as many real challenges for IoT adoption. From a macro perspective, IoT adoption needs to cater to the profound societal and economic changes caused by COVID-19. Since the beginning of the global pandemic,  behaviors of individuals, communities and organisations have undergone a major shift~\cite{5WaysPsy84}. Moreover, some of this shift will not be restored in the next few years or even beyond that. The focus of social construction is on well-being during the pandemic and in post-pandemic world. In the context of financial constraints, IoT adoption must be more effective, efficient, purposeful, and have a significant ROI. 

From a micro perspective, IoT technology requires faster and more revolutionary innovation to ensure the functioning of society, promote civil construction, and respond to possible crises in the future. The global IoT market was growing before the pandemic, but some of its shortcomings have been magnified during the pandemic such as the ease and economy of device installation and data security. Also, new specific problems are emerging in multiple IoT sectors and effective solutions for these problems will be the key in accelerated IoT development and adoption in these sectors.

In this section, we discuss challenges as well as key research directions across different IoT verticals. Specifically, in \autoref{ssec:ubi_challenges}, we focus on general challenges that are ubiquitous to the whole IoT industry. From \autoref{ssec:health_challenges} to \autoref{ssec:iiot_challenges}, we go deeper and summarize the existing and new challenges specific to each IoT vertical. We also discuss key research directions important to facilitate and accelerate IoT adoption.

\subsection{Challenges Ubiquitous to All Verticals}\label{ssec:ubi_challenges}

\subsubsection{Financial Constraints}
A large number of companies,  organisations and individuals are facing financial stress which is one key factor negatively affecting IoT adoption. 
Many organisations have reduced or stopped altogether their investment on many new or planned initiatives such as IoT projects. Another financial challenge caused by COVID-19 is the increasing labor costs for device installation under social contact restrictions.

A key challenge that needs attention is to reduce the cost of IoT system development, installation, and usage. Convergence of IoT and cloud services is important, making the IoT infrastructure that is heavily related to geographic space a virtualized and sustainable resource. Furthermore, open source IoT data processing software, analytics tools and testbeds need to be developed to enable organisations to avoid/reduce the cost of outsourcing or developing these in-house.

It has become ever more important to develop IoT sensors and devices that are cheap and easier to install and  maintain.
For example, it is important to invent devices that operate on minimal power~\cite{Gulati:2018:CSC:3279953.3264920} as this can reduce the maintenance cost, i.e., replacing batteries. One such example is the Ultra Low Power SoC for battery powered IoT devices~\cite{DA16200}, developed by Dialog Semiconductor, which can provide more than one year battery life for many IoT applications.  
Also, it is critical to develop cheap plug-and-play sensing devices~\cite{ravichandran2015wibreathe, Zhao:2014:SDI:2642918.2647380} that can be easily integrated into ordinary facilities without additional hardware investment and deployment cost. 
It will also be helpful to explore self-configuration and self-adaptation of IoT devices~\cite{chatzigiannakis2012true} as well as intelligent human-computer interaction~\cite{chhikara2020federated} to ease the installation and usage of IoT systems.

\subsubsection{Data Security and Privacy}
Many countries have  introduced various emergency regulations during the pandemic such as limits on people's movements, social distance requirements and less strict privacy obligations~\cite{do2020covid}. Once the pandemic is over, such regulations will need to be carefully reviewed to ensure people's rights and privacy are protected~\cite{OECD}. In order to be  better prepared for future pandemics and emergencies, the research community needs to work on robust data access policies, security protocols and privacy-preserving 
solutions for tracking, monitoring and analytics etc.
Also, instead of conventional data centers, decentralized data pools at personal devices may be considered to give more control to the end users over their own data. IoT sensors do not actively collect and upload data but transfer the created data to user devices. Users can choose to distribute data peer-to-peer to their trusted users or third-party applications for data analysis and decision-making. Moreover, lightweight but reliable encryption and network transmission technology are needed for personal devices. The solution should be efficiently integrated with emerging technologies such as 5G, edge computing, and Blockchain~\cite{fan2017blockchain}.

\subsection{Healthcare} \label{ssec:health_challenges}

COVID-19 has made it even more important for healthcare to be ubiquitous, low threshold, and more efficient in monitoring, testing and diagnosis. 
This brings about many new challenges.
First, ubiquity means that IoT-based medical services must be available on-demand and accessible to a wider population. However, most people do not own professional medical equipment, whereas more affordable devices such as smartphones and wearables have limited capabilities in healthcare.
Second, the low threshold requires reducing the cost and difficulty of using IoT-based medical services. However, medical service processes are typically complicated and expensive. Moreover, many people find it hard to adopt and use new technologies such as IoT systems for healthcare. 
Third, it is important to address energy requirements and computational efficiency of healthcare sensors, e.g.,  to enable continuous monitoring and high-quality testing/diagnosis.
Based on these challenges, below we enumerate some promising R\&D directions for healthcare IoT.

\subsubsection{Upgraded Wearable Technology} 

There is a need to make wearables more comfortable to wear/carry. Some recent studies~\cite{metcalf2016wearables, hall2019wearables} have pointed out that the product design of wearables should be optimized for different groups with different lifestyles and circumstances.
For example, to adapt to the needs of patients with respiratory diseases, \citet{ruth_multichannel_2020} incorporate an optical sensor array into a wearable face mask for continuous cardiovascular monitoring.
There is also a need to develop energy harvesting sensors for wearables since smart algorithms on wearables can drain its battery quickly. Therefore, energy harvesting technology utilizing thermometric principle can be applied to wearables, e.g., \citet{IEEEhowto:SelfChargingWearable} demonstrate the feasibility of operating a commercial glucose sensor using only the body heat as its energy source. 

\subsubsection{AIoT (AI + IoT) for Healthcare} 

Algorithms currently running on off-the-shelf IoT devices (e.g., smartphones and wearables) must be able to provide quality comparable to medical grade equipment~\cite{IEEEhowto:ComparisonHRVECG}. In the recent years, AI for healthcare has achieved remarkable progress in screening, disease diagnosis, and telemedicine, using machine learning models driven by massive health data~\cite{lim2020applications}. It is important to accelerate the integration of IoT and AI for healthcare. However, one  urgent issue to solve is the model training and deployment on IoT devices that have limited computational and memory resources. There are preliminary works in this direction. \citet{di2020saia} discuss the split AI architecture for mobile healthcare systems, where lightweight AI solutions can work locally on the client side. \citet{liu2020multi} present a multi-task convolutional network on an ARM CPU to enable real-time cardiovascular and respiratory measurement based on mobile videos. However, there are preliminary efforts and more work is needed in this direction.

\subsubsection{Making Health Services More Accessible} 

People may be unwilling or unable to invest in specialized medical devices or wearables. To make healthcare services accessible to a wider population, a multi-faceted approach is needed. For example, there is a need to establish community-based unmanned clinics~\cite{Unmanned59}. Similarly, it is important to develop low-cost and self-help medical solutions that work on off-the-shelf devices such as smartphones and home audio systems. 
A recent survey~\cite{ueafuea2020potential} found that most existing studies rely on expensive research-grade wearable devices and there is a need to replace these research-grade wearables with the current smartwatches.
Some recent breakthroughs in making health services more accessible include COVID-19 screening using cough recording~\cite{laguarta2020covid}, smartphone-based capture and interpretation of Rapid Diagnostic Tests~\cite{park2020supporting}, and on-device vitals measurement~\cite{liu2020multi}.

\subsection{Smart Homes} \label{ssec:home_challenges}

One of the biggest obstacles in smart homes IoT adoption is the cost and technical skills required for home automation. This requires data generated and collected from disparate sources to be effectively integrated and analyzed.  Therefore, currently, mostly hobbyists and tech enthusiasts invest time and efforts in smart homes. Reducing the setup threshold is critical in facilitating and accelerating user adoption. Thus, there is a need to develop plug-and-play solutions that can be easily deployed and used by people who are not necessarily tech-savvy. 
Also, most of the current smart home devices/services are driven by simple rules such as IFTTT (if-this-then-that), which lacks flexibility and can hardly meet the needs of users considering their daily workloads.
To go beyond the simple rules, there is a need to equip home devices with intelligence and context-awareness that can define rules and operations with no or little human intervention.

Energy sector needs to adapt to the changes brought to our lifestyles and energy consumption patterns by COVID-19. People are spending more time at home and there is a shift in energy consumption from commercial buildings/offices to homes and this trend is expected to stay. 
Energy supply and billing should be optimized based on the occupancy profiles in smart homes and smart buildings in general. 
Considering that energy consumption is a stochastic dynamic process, techniques such as time series analysis and reinforcement learning will help optimize the sustainability of energy supply. Also, efforts should be made to maximize the usage of  renewable energy resources especially on the edge IoT devices.

\subsection{Smart Buildings} \label{ssec:building_challenges}

Smart buildings are an integral part of infection control in the wake of COVID-19. Disinfection and contactless operations implemented by robots and smart devices promise to address a long-standing necessity. However, several challenges need to be addressed.
First, a building compacts together many different functional areas and objects in a relatively small space. They form complex pathways and space usage rules. 
The automation of robots and smart devices must grasp spatial indoor topology (e.g., connectivity and accessibility), semantics (e.g., occupation and usage), and dynamic information (e.g., flow, density, and air quality) of these areas, to better serve people in the building.
Second, privacy and access policies place strict requirements on sensor deployment in buildings. This is exacerbated by disparate data created by different types of sensors that must be integrated for analysis and decision making. 
We highlight two interesting directions for smart buildings in the presence of COVID-19.

\subsubsection{Semantic-aware and Dynamics-aware Modeling} 

It is critical for the effective operation of a smart building to properly capture and represent the semantics and dynamics of the building. Semantic-aware~\cite{feng2020indoor} and dynamics-aware~\cite{misra2020s} models and indexes are required to facilitate data analysis and decision-making processes in smart buildings. For example, using crowd and occupancy information, effective and safe indoor navigation services can be provided to users to avoid contact with other people while navigating indoor spaces~\cite{liu2021towards}. Similarly, analytics techniques must be developed that use indoor mobility data collected from various IoT sensors to understand how indoor spaces are used. This is critical for important smart building operations such as studying potential impact of occupancy and movement on virus transmission and identifying the areas that need frequent cleaning.

\subsubsection{Sensor Data Integration for Context-Awareness} 

It is important to equip robots and other smart devices with context-awareness. This requires data generated and collected from disparate sources to be effectively integrated and analyzed. For example, disinfection robots can be made to utilize data from other in-building sources such as CCTV, swipe-cards access database, temperature sensors, and location data from mobile devices, to identify and clean the areas that were visited by people with elevated temperatures. Similarly, smart HVAC systems can utilize the data obtained from the visitor management system to optimize ventilation and temperature management such that not only the indoor air quality is improved but the power usage is also reduced.

\subsection{Smart Cities} \label{ssec:city_challenges}

To better prepare for future public health emergencies, smart cities need to embrace IoT and other emerging technologies for critical tasks such as social distance measurement, viral testing and infection case reporting~\cite{gupta2020enabling}. 
Governments typically have data from many different departments and sources that are essentially stored and managed in different formats and systems. One critical challenge is to efficiently and effectively integrate data from different sources for data analysis while guaranteeing proper access control. 
Several future directions exist for efficient data integration and integrity~\cite{raghavan2020data}. 
First, it is important to study the open standards to encode IoT sensory data for better interoperability.
Second, it is useful to design smart city ontology representing particular domain knowledge to enrich context information to IoT sensory data before data exploitation.
Third, it is helpful to formulate data sharing and integration schemes to facilitate different authorities sharing their sensor data without costly restructuring of existing data architecture.

COVID-19 has given a huge blow to the brick-and-mortar retail and entertainment industries.
People are expected to continue spending more time and money online.
In this case, the services provided by brick-and-mortar stores need to be innovative to attract customers and meet their demands.
The retail and entertainment industry could better integrate AI, IoT, and big data analysis, to achieve a comprehensive understanding of customers, places and products. 
To this end, personalized modeling and graph representation of user interactions are directions worth exploring.

\subsection{Transportation} \label{ssec:transport_challenges}

COVID-19 has significantly impacted the transportation networks, both in urban transport sector (e.g., metro, bus, taxi) and long-distance transport sector (e.g., air travel and train). To avoid person-to-person contact, people's travel patterns (time and route planning), selection and operating mode of pickup sites, social behavior of drivers and passengers, and security checks have all changed. 
IoT adoption to transportation must adapt to these new changes fast, to facilitate the activities of passengers and drivers. This calls for data-driven and self-adaptive decision-making processes over IoT data.
An urgent need is to use the tracking and monitoring capabilities of IoT to share data with the travelers. Travelers are often eager to obtain information regarding their upcoming travel, e.g., the number of passengers in the cabin, the waiting time to pass the security and viral checks, and the number of positive cases detected recently along their travel routes.
A key challenge is to seamlessly integrate data from different sources while ensuring the accuracy and reliability of the information demanded~\cite{darsena2020safe, misra2020s}.

\end{paracol}
\nointerlineskip
\begin{specialtable}[htbp] 
\caption{A summary of major challenges and key research directions. \label{table:challenges}}
\begin{tabular*}{\textwidth}{p{35mm}p{140mm}}
\toprule
\textbf{Sector}	& \textbf{Challenges and key research directions}	\\
\midrule
\textbf{All} & reducing the cost of developing, installing and using IoT solutions and systems, data security and privacy\\ \\ 
\textbf{Healthcare} & low-energy or energy harvesting wearables, research-grade wearables, AI algorithms on healthcare devices (edge AI), more-accessible healthcare \\ \\ 
\textbf{Smart Homes} & designing cheap and easy-to-use IoT devices, more customisable and intelligent IoT devices, smart energy management \\ \\ 
\textbf{Smart Buildings} & semantic-aware and dynamics-aware indoor modeling, sensor data integration and exploitation, context-awareness\\ \\ 
\textbf{Smart Cities} & integrating  and sharing data from disparate sources, standardized smart city architectures, personalized modeling\\ \\ 
\textbf{Transportation} & data-driven and self-adaptive decision making, generating IoT tracking \& monitoring data and effective techniques to share it with travellers, traceable and transparent logistics, vaccine passports\\ \\
\textbf{Industrial IoT} & virtual reality based remote work, digital twins\\
\bottomrule
\end{tabular*}
\end{specialtable}
\begin{paracol}{2}
\switchcolumn

It is also critical to improve the logistics to avoid delays, minimize social contact, and hoarding of supplies useful in dealing with the pandemic. This requires rethinking of previous tenets and ontologies in supply chain~\cite{sarkis2020supply}. To improve the efficiency of logistics, it is necessary to consider and prioritize the tasks in supply chain, capabilities and preferences of carriers, and  real-time traffic that has shown the prospects of combining spatial crowdsourcing and personalized modeling~\cite{ruan2020doing, ding2020delivery}.
Moreover, logistics should be made fully traceable and transparent.
Current promising research directions include the development of sustainable tracking tags and ultra-low-power tracing solutions~\cite{perez2020fast}, and the adoption of blockchain to achieve trustworthy and long-term preservation of product histories~\cite{francisco2018supply}.

\end{paracol}
\nointerlineskip
\begin{specialtable}[H]
\widetable
\caption{Our prediction of the COVID-19's impact on IoT adoption in short-term and mid- to long-term. \p{} indicates accelerated adoption, \n{} indicates decelerated adoption, and \co{} indicates that the adoption is both positively and negatively impacted in different subareas. Number of \p{} and \n{} indicate how strong the positive and negative impacts are expected to be, respectively.\label{table:summary}}
\setlength{\tabcolsep}{1pt} 
\renewcommand{\arraystretch}{0.8}
\fontsize{9}{9}\selectfont\begin{tabular*}{\textwidth}{@{} lccccccp{68.6mm} @{} }
        \toprule
        {\textbf{Sector}} & \multicolumn{3}{c}{\textbf{Short-Term}} & \multicolumn{3}{c}{\textbf{Mid- to Long-Term}} & {\textbf{Explanation}} \\
        \cmidrule{2-7}
        &\textbf{ Demand} &\textbf{ Investment }& \textbf{Adoption} &\textbf{ Demand }&\textbf{ Investment }& \textbf{Adoption }& \\
        \midrule
        \multirow{9}{*}{Healthcare} & \multirow{9}{*}{\p\p\p} & \multirow{9}{*}{\p\p\p} & \multirow{9}{*}{\p\p\p} & \multirow{9}{*}{\p\p\p} & \multirow{9}{*}{\p\p\p} & \multirow{9}{*}{\p\p\p} & There is a huge demand for efficient and effective healthcare driven by larger number of patients, overburdened facilities and overworked hospital staff due to the COVID-19. Governments and organisations have invested heavily to combat the pandemic and IoT adoption is expected to grow at a significant pace both in the short-term and mid- to long-term.\\ \\
        \multirow{7}{*}{Smart Homes} & \multirow{7}{*}{\p} & \multirow{7}{*}{\n} & \multirow{7}{*}{\n} & \multirow{7}{*}{\p} & \multirow{7}{*}{\p} & \multirow{7}{*}{\p} & People are spending more and more time at home due to lockdowns and work/study from home trends that are expected to stay. Financial stress on households is holding back the IoT adoption in the short-term but an accelerated adoption is expected in the mid- to long-term. \\ \\
        \multirow{10}{*}{Smart Buildings} & \multirow{10}{*}{\p} & \multirow{10}{*}{\n\n} & \multirow{10}{*}{\n} & \multirow{10}{*}{\p} & \multirow{10}{*}{\p} & \multirow{10}{*}{\p} & Tracking, monitoring and cleaning requirements imposed by governments have increased the demand for IoT. However, the recession has made it difficult for most of the organisations to invest in new IoT projects. Thus, despite some initiatives, the short-term impact is negative on IoT adoption. However, in the mid- to long-term, accelerated adoption is expected once the financial stress is~eased.  \\ \\
        \multirow{7}{*}{Smart Cities} & \multirow{7}{*}{\p\p} & \multirow{7}{*}{\p\p} & \multirow{7}{*}{\p\p} & \multirow{7}{*}{\p\p} & \multirow{7}{*}{\p\p\p} & \multirow{7}{*}{\p\p\p} & Smart and resilient cities have become even more important to be prepared for future pandemics and emergencies. IoT is being deployed for various applications especially tracking and monitoring. Adoption is expected to gain pace in mid- to long- term when financial stress is eased. \\ \\
         \multirow{10}{*}{Transportation} & \multirow{10}{*}{\n} & \multirow{10}{*}{\n\n} &  \multirow{10}{*}{\n} & \multirow{10}{*}{\p} & \multirow{10}{*}{\co} & \multirow{10}{*}{\co} & Sharp decline in transportation usage has adversely affected revenue, which has negatively affected the adoption of IoT. Although some areas (e.g., logistics, in-cabin analytics) are still gaining some attention, overall trend is negative. Mid- to long-term impact appears to be mixed with accelerated adoption expected in logistics but relatively stable in public and private transports. \\ \\
        \multirow{9}{*}{Industrial IoT} & \multirow{9}{*}{\p\p} & \multirow{9}{*}{\n\n} &  \multirow{9}{*}{\n\n} & \multirow{9}{*}{\p\p} & \multirow{9}{*}{\p\p} & \multirow{9}{*}{\p\p} & Applications like remote equipment monitoring and inventory management have increased the demand but the industries are finding it hard to invest in IoT due to the financial strain. In the mid- to long-term, an accelerated growth is expected due to the importance of IoT to keep industries operational in future pandemics or emergencies. \\ 
        \bottomrule
   \end{tabular*}
\end{specialtable}
\begin{paracol}{2}
\switchcolumn

\subsection{Industrial IoT} \label{ssec:iiot_challenges}

COVID-19 has disrupted industries and IoT has become even  more important for industries to ensure business continuity, employee safety, remote asset control, and remote collaboration etc. To ensure that industries are ready for future pandemics, research and development into various important directions are needed such as digital maintenance of equipment, end-to-end automation, advanced remote collaboration and asset control etc. Virtual reality is becoming more important to allow workers effectively operate equipment remotely. Digital twins~\cite{magargle2017simulation} are also essential for industries to adapt to the new normal, e.g., to identify potential equipment failures and to automate operations.

\change{
\autoref{table:challenges} provides a summary of major challenges and key research directions discussed in this section.}


\section{Conclusions}\label{sec:conclusion}

COVID-19 is proving to be a catalyst for technology adoption and innovation. We have thoroughly analyzed recent research literature, examined reports from leading consulting firms and  interviewed many experts in different sectors.  Based on this, we discuss the potential impact of COVID-19 on IoT adoption in different sectors namely healthcare, transportation, industrial IoT, smart homes, smart buildings and smart cities. 
We also discuss different initiatives being undertaken  in different sectors in the wake of this pandemic. Furthermore, we highlight various challenges that must be addressed and important research directions that need attention to facilitate accelerated IoT adoption in different sectors.
In \autoref{table:initiatives}, we summarized some of the most notable recent initiatives taken in wake of COVID-19. \autoref{table:challenges} provides a summary of major challenges and key research directions to facilitate IoT adoption in different sectors. \autoref{table:summary} presents a summary of our predicted short-term and mid- to long-term impact across these sectors.

\funding{This research is partially supported by the Australian Research Council (ARC) FT180100140.}





\end{paracol}

\vspace{6pt} 

\reftitle{References}

\section*{Short Biography of Authors}
\bio
{\raisebox{-0.35cm}{\includegraphics[width=3.5cm,height=5.3cm,clip,keepaspectratio]{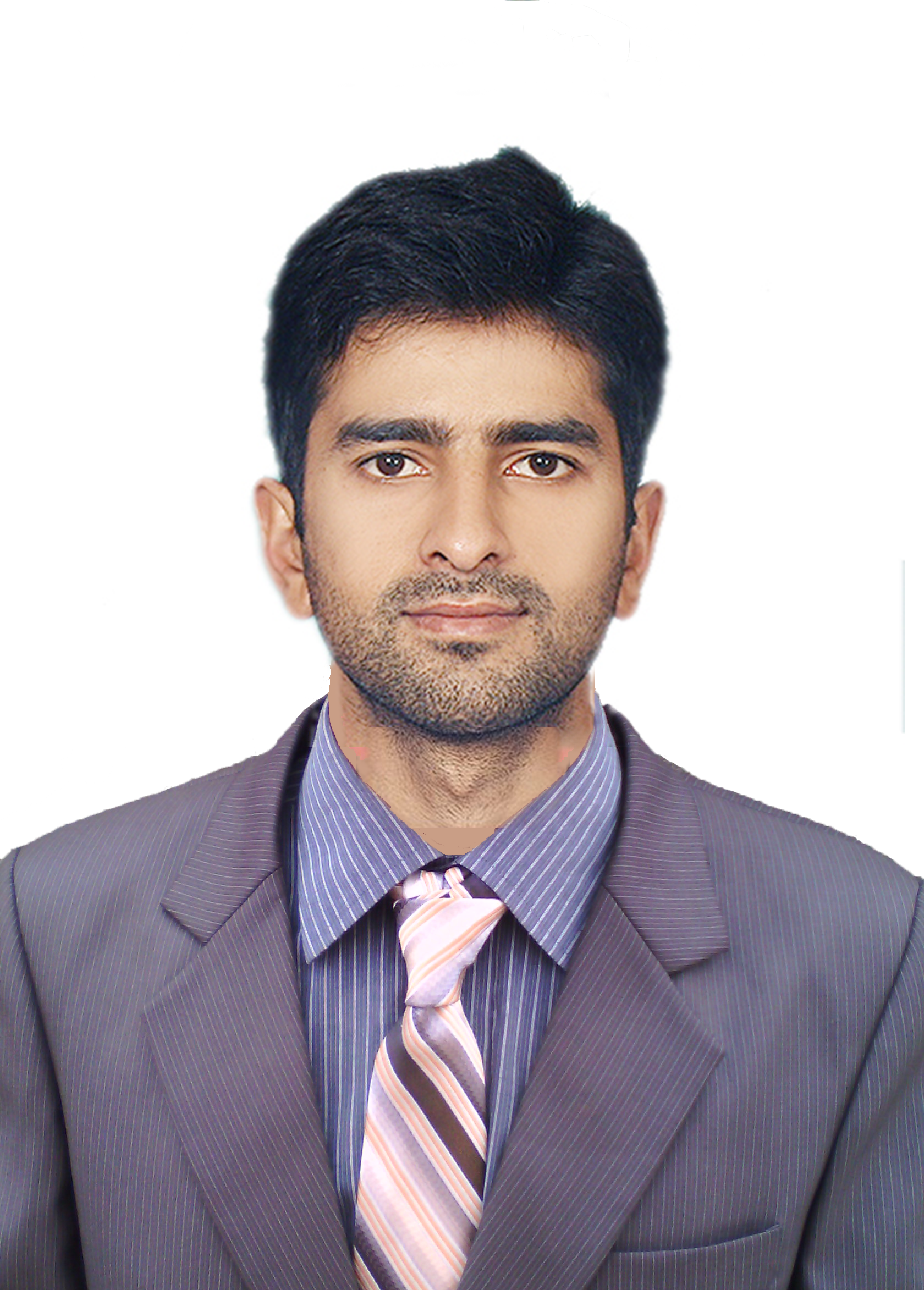}}}
{\textbf{Muhammad Umair} is a Lecturer at Department of Electrical,
Electronics and Telecommunication Engineering, New Campus, UET Lahore. He completed his B.Sc. Electrical Engineering and M.Sc. Electrical Engineering from University of Engineering \& Technology (UET) Lahore in 2014 and 2017, respectively. He has worked as a Research Officer at Internet of Things (IoT) lab at Al-Khwarizmi Institute of Computer Sciences, UET Lahore. He has also worked at Sultan Qaboos IT Research lab as a Research Officer. His survey on Social IoT platforms is the most cited survey for SIoT applications. He has designed graduate level courses on IoT. His areas of interests include Internet of Things, Embedded Systems, Network Systems, Machine Learning, Algorithms Development, Pervasive Computing, Ubiquitous Computing, Cloud Based Systems, Data Analytics and working on application layer of any of the defined problems.}
\bio
{\raisebox{-0.35cm}{\includegraphics[width=3.5cm,height=5.3cm,clip,keepaspectratio]{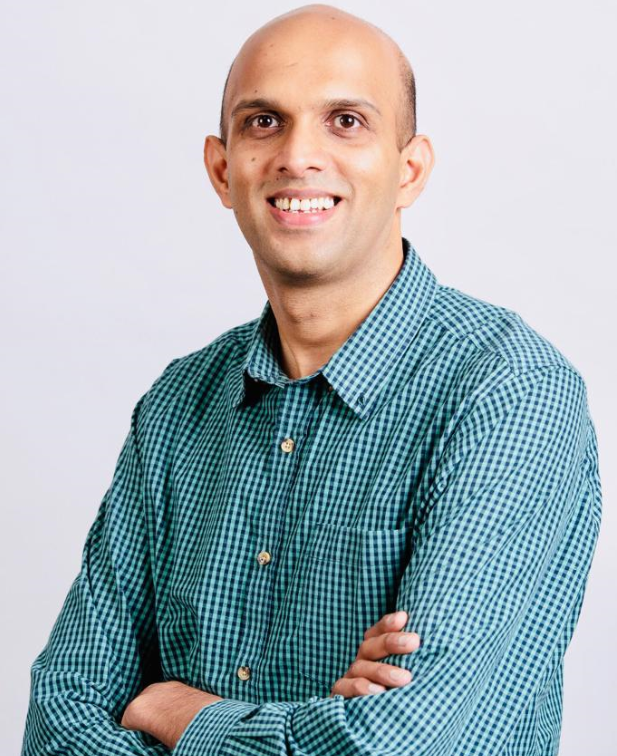}}}
{\textbf{Muhammad Aamir Cheema} is an ARC Future Fellow, an Associate Professor and Director of Research at the Department of Software Systems and Cybersecurity, Faculty of Information Technology,  Monash University, Australia. He obtained his PhD from UNSW Australia in 2011. He is the recipient of 2012 Malcolm Chaikin Prize for Research Excellence in Engineering, 2013 Discovery Early Career Researcher Award, 2014 Dean’s Award for Excellence in Research by an Early Career Researcher, 2018 Future Fellowship, 2018 Monash Student Association Teaching Award and 2019 Young Tall Poppy Science Award. He has also won two CiSRA best research paper of the year awards, two invited papers in the special issue of IEEE TKDE on the best papers of ICDE, and three best paper awards at ICAPS 2020, WISE 2013 and ADC 2010, respectively. He is the Associate Editor of IEEE TKDE and DAPD and served as PC co-chair for ADC 2015, ADC 2016, 8th ACM SIGSPATIAL Workshop ISA 2016 \& 2018, IWSC 2017, proceedings chair for DASFAA 2015 \& ICDE 2019.}

\bio
{\raisebox{-0.35cm}{\includegraphics[width=3.5cm,height=5.3cm,clip,keepaspectratio]{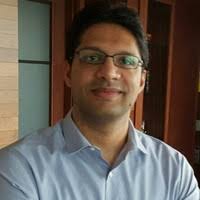}}}
{\textbf{Omer Cheema} is the head of IoT Wi-Fi business unit at Dialog Semiconductor. Prior to Dialog, Dr. Cheema was working at Samsung leading marketing teams of its IoT business. Dr. Cheema is a technology veteran with about 20 years experience in corporate strategy, marketing, management consulting and technology development.  Earlier in his career, he worked as an ASIC design engineer at AMD.  Dr. Cheema holds a PhD in semiconductor design from ENSTA Paris and MBA from INSEAD.}

\bio
{\raisebox{-0.35cm}{\includegraphics[width=3.5cm,height=5.3cm,clip,keepaspectratio]{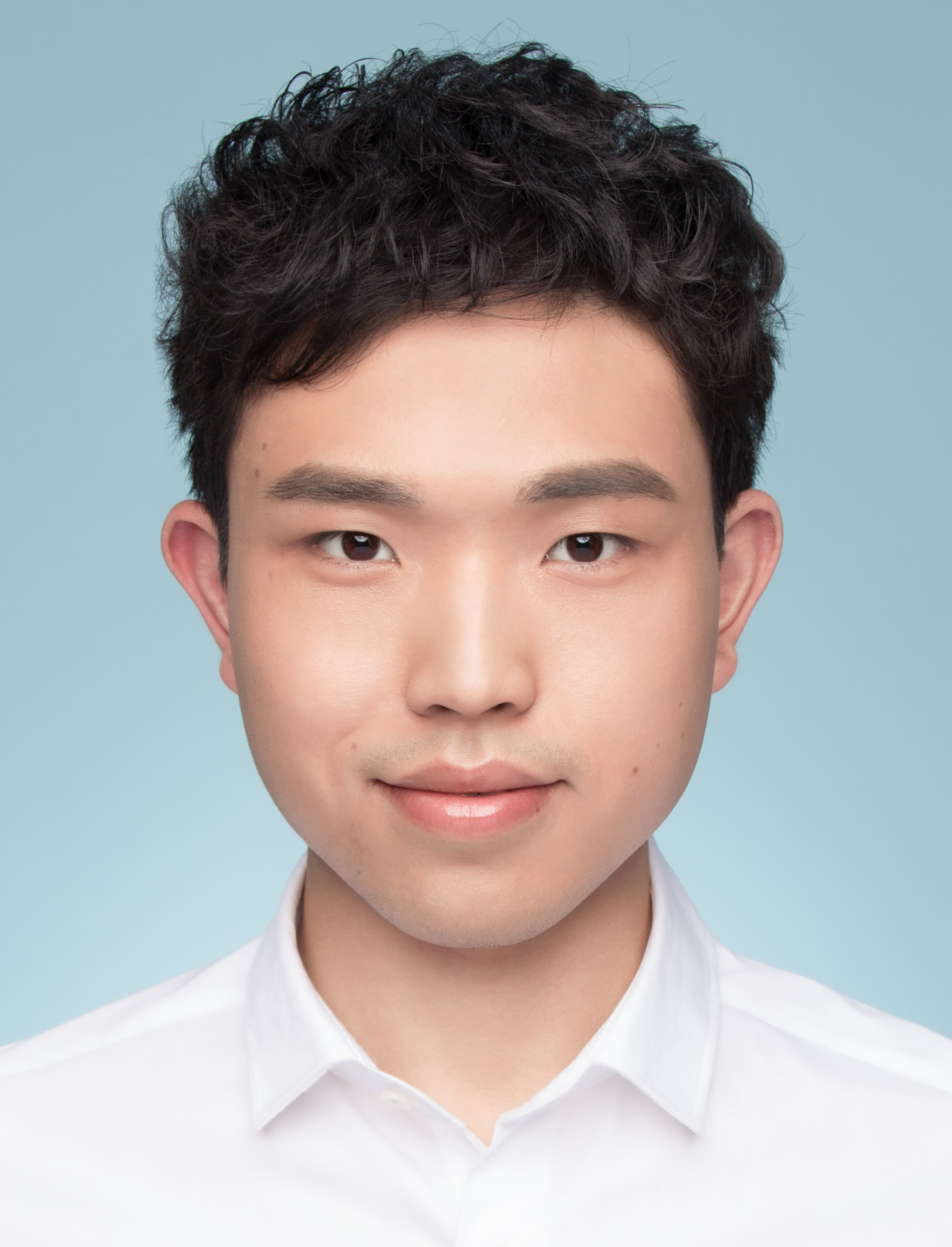}}}
{\textbf{Huan Li} is an Assistant Professor and the EU Marie Curie IF Fellow with the Department of Computer Science, Aalborg University, Denmark.
He received the PhD degree in computer science from Zhejiang University, China in 2018. His research interests include IoT data management, spatio-temporal data analytics, and mobile/pervasive computing. He is a member of the IEEE and the ACM.}

\bio
{\raisebox{-0.35cm}{\includegraphics[width=3.5cm,height=5.3cm,clip,keepaspectratio]{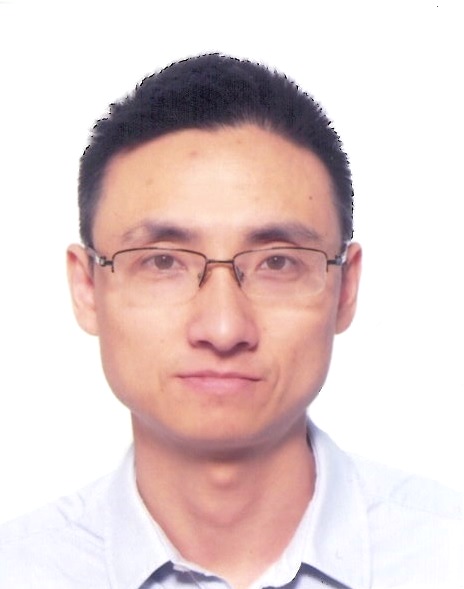}}}
{\textbf{Hua Lu} is a Professor of Computer Science in the Department of People and Technology, Roskilde University, Denmark. He received the BSc and MSc degrees from Peking University, China, and the PhD degree in computer science from National University of Singapore. His research interests include database and data management, geographic information systems, and mobile computing. He has served as PC cochair or vice chair for ISA 2011, MUE 2011, MDM 2012 and NDBC 2019, demo chair for SSDBM 2014, and PhD forum cochair for MDM 2016. He has served on the program committees for conferences such as VLDB, ICDE, KDD, WWW, CIKM, DASFAA, ACM SIGSPATIAL, SSTD, MDM, PAKDD, and APWeb. He received the Best Vision Paper Award at SSTD 2019. He is a senior member of the IEEE.}


\end{document}